\documentclass[]{aa}  

\DeclareUnicodeCharacter{2212}{-} 
\usepackage{booktabs}
\usepackage{siunitx}
\usepackage{graphicx}
\usepackage{multirow}
\usepackage{gensymb}
\usepackage{caption}
\usepackage{subcaption}

\usepackage{txfonts}
\usepackage{hyperref}
\begin{document}

   \title{Quasar and galaxy classification using Gaia EDR3 and CatWise2020}

\author{Arvind C.N. Hughes
          \inst{1,2}
          \and
          Coryn A.L.\ Bailer-Jones
          \inst{1}
          \and
          Sara Jamal
          \inst{1}
          }

   \institute{Max Planck Institute for Astronomy, K\"onigstuhl 17, 69117 Heidelberg, Germany\\
              \email{ahughes@mpia.de}
         \and
          School of Mathematical and Physical Sciences, Macquarie University, Sydney, NSW 2109, Australia
          \and
           Research Centre in Astronomy, Astrophysics \& Astrophotonics, Macquarie University, Sydney, NSW 2109, Australia 
             }

   \date{Received 1 September 2022 / Revised 13 September 2022 / Accepted 8 October 2022} 

\abstract {
In this work, we assess the combined use of Gaia photometry and astrometry with infrared data from CatWISE in improving the identification of extragalactic sources compared to the classification obtained using Gaia data. Here we perform a comprehensive study in which we assess different input feature configurations and prior functions to identify extragalactic sources in Gaia, with the aim of presenting a classification methodology integrating prior knowledge stemming from realistic class distributions in the universe.
In our work, we compare different classifiers, namely Gaussian Mixture Models (GMMs) and the boosted decision trees, XGBoost and CatBoost, in a supervised approach, and classify sources into three classes - star, quasar, and galaxy, with the target quasar and galaxy class labels obtained from the Sloan Digital Sky Survey Data Release 16 (SDSS16) and the star label from Gaia EDR3. 
In our approach, we adjust the posterior probabilities to reflect the intrinsic distribution of extragalactic sources in the universe via a prior function. In particular, we introduce two priors, a global prior reflecting the overall rarity of quasars and galaxies, and a mixed prior that incorporates in addition the distribution of the extragalactic sources as a function of Galactic latitude and magnitude. 
   Our best classification performances, in terms of completeness and purity of the extragalactic classes, namely the galaxy and quasar classes, are achieved using the mixed prior for sources at high latitudes and in the magnitude range G = 18.5 to 19.5. 
   We apply our identified best-performing classifier to three application datasets from Gaia Data Release 3 (GDR3), and find that the global prior is more conservative in what it considers to be a quasar or a galaxy compared to the mixed prior. 
   In particular, when applied to the quasar and galaxy candidates tables from GDR3, the classifier using a global prior achieves purities of 55\% for quasars and 93\% for galaxies, and purities of 59\% and 91\% respectively using the mixed prior.
   When compared to the  performances obtained on the GDR3 pure quasar and galaxy candidates samples, we reach a higher level of purity, 97\% for quasars and 99.9\% for galaxies using the global prior, and purities of 96\% and 99\% respectively using the mixed prior. 
   When refining the GDR3 candidates tables via a cross-match with SDSS DR16 confirmed quasars and galaxies, the classifier reaches purities of 99.8\% for quasars and 99.9\% for galaxies using a global prior, and 99.9\% and 99.9\% using the mixed prior.
   We conclude our work by discussing the importance of applying adjusted priors portraying realistic class distributions in the universe and the effect of introducing infrared data as ancillary inputs in the identification of extragalactic sources.

   }

   \maketitle
%
\section{Introduction}
Classification of galactic and extragalactic sources is fundamental for statistical analyses of large populations, as well as probing the properties of individual objects. For instance, quasars (quasi-stellar objects) refer to highly luminous active galactic nuclei used as probes to investigate fundamental questions in Cosmology such as galaxy evolution \citep[e.g.,][]{harrison_agn_2018}, the composition of the interstellar medium \citep[e.g.,][]{li_spatially_2022} and supermassive black-hole formation and evolution \citep[e.g.,][]{croom_quasar_2009}.

All-sky surveys, such as the Sloan-Digital Sky Survey (SDSS) \citep{york_sloan_2000} and the Wide-field Infrared Survey Explorer (WISE) \citep{wright_wide-field_2010}, have created detailed maps of the universe at optical and infrared wavelengths. Infrared data is highly informative for the classification of stars, quasars and galaxies. As demonstrated in the work by \cite{kurcz_towards_2016}, the authors exploited the infrared colours from WISE and reported a 90--95\% classification accuracy across all object types despite the limitations observed for galaxy sources with a high dust component. The combined use of infrared data with optical photometry should, in principle, enhance the classification accuracy and reduce the number of false positives  across all object types.

However, a large fraction of work on classification fails to consider the intrinsic distribution of sources of different classes, and only reports results, in particular the accuracy (i.e., the fraction of correct predictions per target class), using a test set that is typically not representative of the observable universe. Moreover, such test sets often under-represent the stellar contaminants that would, in practice, lower the purity of extragalactic classification. To account for the actual distribution, we introduce a prior (discussed in detail in Sect.~\ref{Subsec:The prior}) which, in a Bayesian framework, is used to adjust the estimated model posterior probabilities in order to reflect the class distribution of sources we would expect to exist. Furthermore, after a model has been applied, we apply an adjustment factor to the distribution of sources, such that the performance metrics are computed as if the model had been applied to the dataset with a realistic expected distribution. Applying both the prior and the adjustment factor result in classification performances that are more representative of what we can achieve -- although inevitably lower -- compared to the results obtained when the prior and adjustment factor had not been applied. Despite the lower results of some models, applying the prior correction is a necessary step because it would depict real classification performances especially for large-scale surveys for which the observed sources are unknown.

The Gaia mission is an optical mapping survey designed to focus on stars in our Galaxy \citep{gaia_collaboration_gaia_2016}.
During Gaia's scan of the entire sky, the satellite observes all point-source-like objects down to a magnitude limit of $G \simeq 21$, including extragalactic objects \citep{gaia_collaboration_gaia_2022}. A reliable methodology to identify extragalactic sources would benefit the construction of comprehensive catalogues useful for addressing fundamental questions in astronomy.

To construct this method, we follow a similar approach to the work by \cite{bailer-jones_quasar_2019}, in which the classification of extragalactic sources was obtained using Gaussian Mixture Models \citep{fraley_model-based_2002} applied to Gaia Data Release 2 photometry and astrometry. In this study, we consider photometry and astrometry from Gaia Early Data Release 3 (GeDR3) and the addition of infrared photometry from CatWISE2020 \citep{marocco_catwise2020_2021}, as well as the application of gradient boosting decision trees, namely XGBoost \citep{chen_xgboost_2016} and CatBoost \citep{dorogush_catboost_2017}, to construct a three-class classifier (quasar, galaxy, star). The objective of our work is to assess the effect of additional information from infrared photometry, the omission of parallax and proper motions on the classification of extragalactic sources, and to evaluate different classification algorithms as well as the appropriate use of different priors to ensure the performance results reported are reflective of reality.
\section{Data}
Our input data comprises astrometry and photometry from the GeDR3 catalogue and infrared photometry from the CatWISE2020 catalogue.  
The training and test datasets for the quasars and galaxies are based on the sixteenth data release of SDSS (SDSS-DR16, \citeauthor{ahumada_16th_2020} \citeyear{ahumada_16th_2020}) while the stars sample is built from the  GeDR3 catalogue. We are aware that SDSS is not complete and does not cover the same magnitudes as Gaia, however we accept these limitations when building our class samples.

The early third release of Gaia data \citep{gaia_collaboration_gaia_2021} was published on 3 December 2020 for observations acquired between 25 July 2014 and 28 May 2017, spanning a period of 34 months. GeDR3 consists of astrometry, and broad band photometry in the G, $\mathrm{G_{\rm BP}}$, and $\mathrm{G_{\rm RP}}$ bands for about 1.8 billion sources. In this work, we set a limit in magnitude up to $G>14.5$\,mag. This work commenced prior to Gaia Data Release 3 (GDR3) and therefore made use of the public data in the early data release. However, since the photometry and astrometry remain unchanged between GeDR3 and GDR3, our work applies to DR3.

The CatWISE2020 catalogue consists of about $1.8$ billion sources observed across the entire sky selected from the WISE and NEOWISE survey data in the W1 and W2 (\SI{3.4}{\micro\metre} and \SI{4.6}{\micro\metre}) bands~\citep{marocco_catwise2020_2021}. In our study, we chose CatWISE2020 instead of All/unWISE as the CatWISE2020 catalogue extends to fainter magnitudes and the associated data processing pipeline uses the full-depth unWISE coaddition of AllWISE and NEOWISE 2019 Data Release for aperture photometry~\citep{marocco_catwise2020_2021}, which results in a significant improvement over the AllWISE data. A 5 arc-second positional cross-match of CatWISE2020 with GeDR3 identifies about $1.5$ billion sources.

\subsection{Classes}\label{sec:classes}
The goal of our classification is to identify objects in the target star, quasar, and galaxy classes. The definition of the target classes is similar to the work by \cite{bailer-jones_quasar_2019}, augmented with the aforementioned CatWISE2020 cross-match. However, in our application we do not use, and thus do not require to be available, the parallax and proper motions. This approach results in a much larger set of galaxies, because most galaxies observed by Gaia lack published parallax and proper motions due to a poor fit of the astrometric model on account of their physical extent. As these may indicate a different type of galaxy, these effectively changes our class definition. We ensure there are no common sources between the three class datasets.

\subsubsection{Quasars}
The SDSS-DR16 quasar catalogue \citep{lyke_sloan_2020} contains 750\,414  quasars confirmed by optical spectroscopy.
Its authors estimate the contamination to be around $0.5\%$. We select objects with a \texttt{zWarning} flag equal to zero, indicating a higher reliability in the classification or the redshift estimation. We cross-match the selected sample to GeDR3 by sky position with a 1 arcsecond search radius using the CDS X-match tool, finding 489\,581 matches in total. 
This constructed sample is afterwards compared with the cross-matched sample from GeDR3 and CatWISE2020 catalogue, resulting in 484\,749 objects with GeDR3 features and CatWISE2020 magnitude measurements in the W1 and W2 bands.

\subsubsection{Galaxies}\label{subsubsec:galaxies}
\begin{figure}[h]
\centering
\includegraphics[scale=0.35]{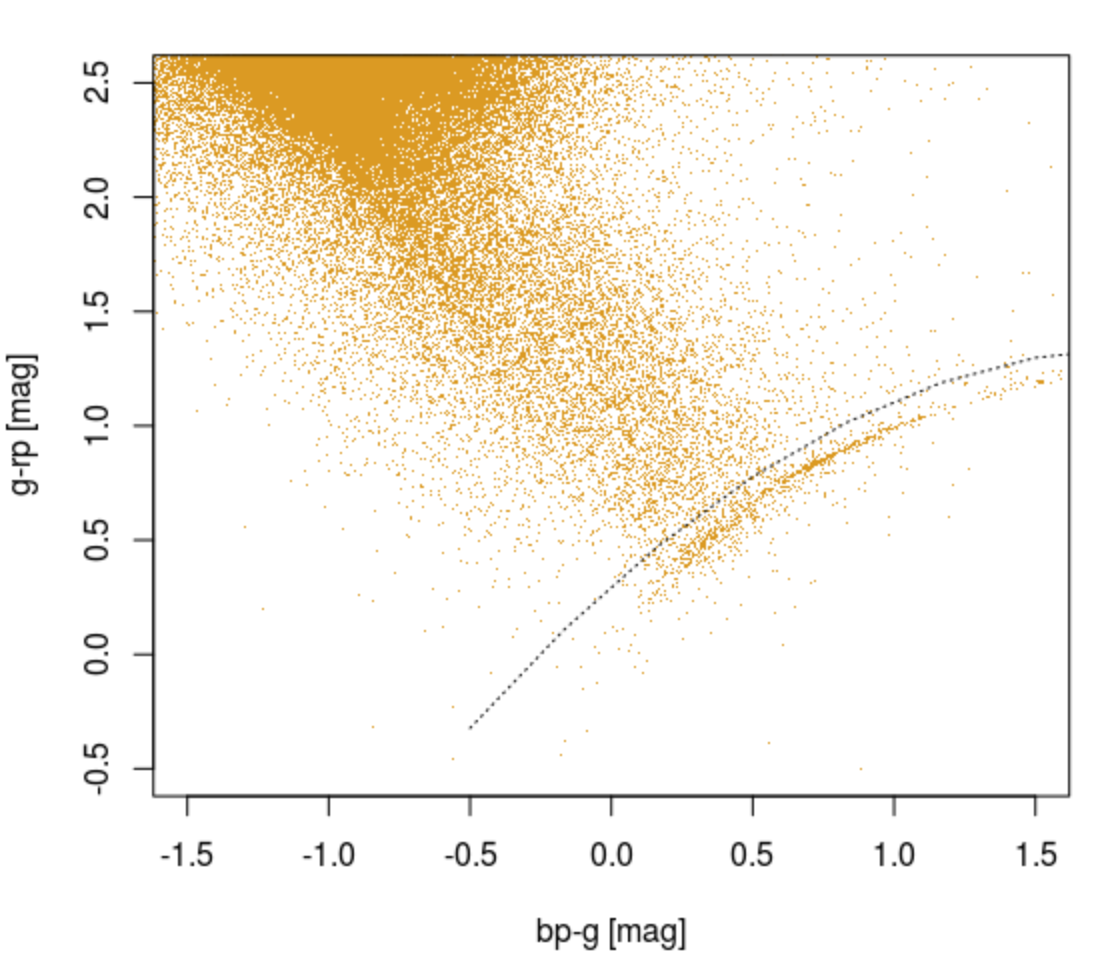}
\caption{Colour-colour diagram of the galaxy class. Sources below the below the dashed-line are contaminants that are removed from the galaxy sample.}
\label{fig:Data-Colour_colour_edge}
\end{figure}

The sample of galaxies in our train and test datasets is constructed from SDSS-DR16~\cite{ahumada_16th_2020, blanton_sloan_2017}. We select 777\,409 objects from the \texttt{SDSS SpecObjAll} table on the SDSS Skyserver identified as \texttt{GALAXY} with \texttt{zWarning} equal to zero, and are identified as neither \texttt{AGN} nor \texttt{AGN BROADLINE} in the subclass field. The selected sample is similarly cross-matched with GeDR3, finding all objects. In our selection, we relax the requirement of the parallax and proper motions, as such information may be unavailable for several sources in Gaia, particularly amongst galaxy sources. Applying the defined criterion, we retain about $90\%$ of the galaxy sources. Furthermore, supplementing the CatWISE2020 colours to our constructed sample results in a total of 766\,310 objects.
Following the work by \cite{bailer-jones_quasar_2019}, we apply a colour-cut to the galaxy sources using the same colour-edge locus as shown in Fig.~\ref{fig:Data-Colour_colour_edge}. Objects below this locus represent stellar contaminants within the galaxy sample. Potential sources of contamination include errors in the SDSS classification or the Gaia BP/RP spectra affected by blends of nearby objects \citep{de_angeli_gaia_2022}. The color cut removes 1061 contaminants from our galaxy sample.

\subsubsection{Stars}
The spectroscopic selection for stars from SDSS data is complex, ill-defined, and likely affected by a biased distribution of stellar types. We therefore do no use SDSS to define the star class.
We exploit the fact that the majority of observed sources in Gaia are expected to be stars. We therefore construct our stars sample via a random subset of 3 million sources from the Gaia catalogue in which known galaxies and quasars are filtered out. We augment the sample with the CatWISE2020 cross-match, resulting in 1.8 million sources identified as stars.
In our constructed stars sample, we could expect a non-zero level of contamination from non-stellar sources. This contamination level is unknown, but our prior defined in Sect.~\ref{Subsec:The prior} is our expectation. Ideally, our classifier trained on the cleaned sample would be robust to contamination.

\subsection{Training, validation, and test sets}\label{Subsec:Training and test}
The full dataset is the combination of the quasar, galaxy, and star samples. The data are split into two equal parts at random. The first part is then split in a ratio 9:1, with the larger used for training, and the latter for validation, i.e. monitoring the performance during the training. For brevity we often refer to these two together as "the training data". The second part is the test set which is kept back to assess the fixed models.

During the training phase, the training dataset is used to train the statistical model while the validation set is used to assess the performance of the trained model. After convergence, the trained model is stored and the test set, i.e. a subset of the data unseen during training, is used to evaluate the performances of the classifier.

For the classifier trained on the balanced dataset, we select a random subset of 200\,000 sources of each class for the training (90\,000 for training and 10\,000 for validation) and test datasets. By constructing a balanced classifier, we are able to directly compare the intrinsic performance of the models trained on different feature configurations and classification methods and identify the best performing method. The class imbalance is addressed in the discussion of the priors in Sect.~\ref{Subsec:The prior}. 

Having selected the best performing model using a balanced training and test dataset discussed in Sect.~\ref{Subsec:Balanced results}, we re-define our training and test datasets to use as many of the available sources in the quasar and galaxy class as possible by sampling a random subset of 900\,000 stars, 200\,000 quasars and 370\,000 galaxies from the full dataset. We train the feature configuration and classification method identified using the balanced dataset on this imbalanced training and test set in Sect.~\ref{Subsec:Max Source Classifier}. This resulting classifier is applied to the application sets in Sect.~\ref{Sec:Application}.
\subsection{Application sets}\label{Sec:Application Datasets}

We use three datasets derived from Gaia Data Release 3 (GDR3)~\cite{vallenari_gaia_2022} to demonstrate the application of our best-performing classifier.

\begin{itemize}
    \item A subset of 50 million randomly selected GDR3 sources that have CatWISE2020 photometry.
    \item Quasar candidate table from GDR3: The quasar tables described in \cite{gaia_collaboration_gaia_2022} represent datasets where there is an estimation of the number of quasars within GDR3. The quasar candidate table contains 6\,649\,162 sources with a purity of 0.52, and is refined further in the pure subsample (1\,942\,825 sources) with a purity of 0.96. 
    \item Galaxy candidate table from GDR3: Similarly defined in \cite{gaia_collaboration_gaia_2022}, the full table reports 4\,842\,342 candidates with a purity of 0.69, and the pure sample (2\,891\,132 sources) reaches a purity of 0.94.
\end{itemize}

There are 144\,109 sources in common between the quasar and galaxy candidates~\cite{gaia_collaboration_gaia_2022}. To make our results on these tables more interpretable we therefore choose to remove these sources from the subsequent analysis.

\subsection{Feature selection}
In feature selection, an important condition is the completeness of each feature, as missing data often cause many statistical methods to fail. As noted in Sect.~\ref{sec:classes}, a large fraction of galaxies do not have published parallaxes and proper motions in GeDR3. We therefore disregard both as input features, in order to retain as many sources as possible.
As inputs to the classifier, we test various combinations of eight features: six of the eight features are defined in ~\cite{bailer-jones_quasar_2019}  which we refer to as "Gaia\_f" and the two features, W1-W2 and the G-W1 colour constructed from the CatWISE2020 catalogue.
The six features from "Gaia\_f": apparent magnitude (G), sine of the Galactic latitude ($sin b$), g-rp (G-RP), bp-g (G-RP), relative variability in the G band ($\rm relvarg$) and the astrometric unit weight error ($\rm UWE$).
As noted in Sect.~\ref{sec:classes} we exclude the parallax and proper motion measurements.
We report the distribution of each feature in Fig.~\ref{fig:Data-Feature_Hists} and their descriptions below:


\begin{figure*}[]
\centering
\includegraphics[scale=0.9]{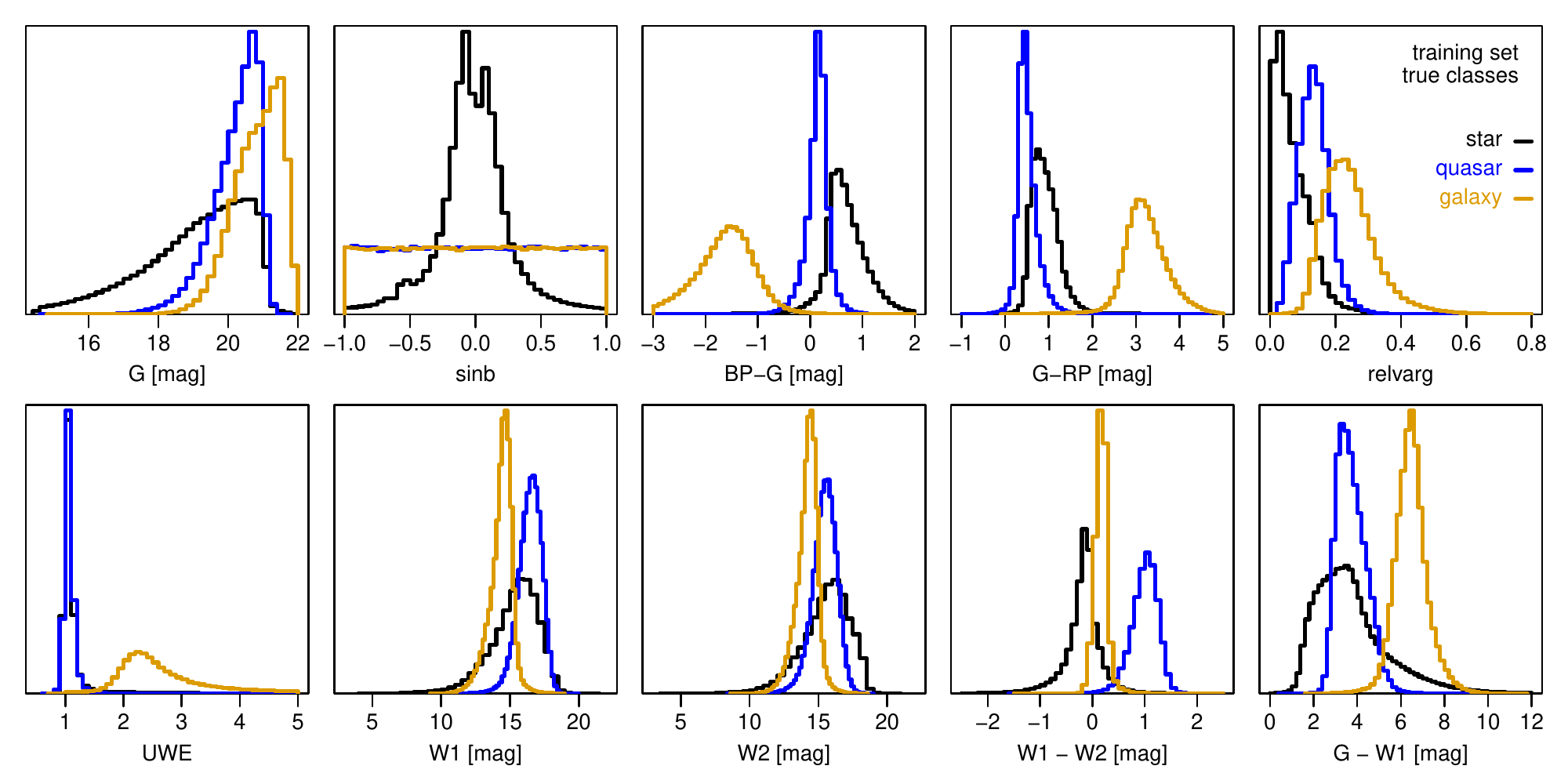}
\caption{Distribution of the features from the training dataset, coloured according to their true classes. Black: stars, blue: quasars, orange: galaxies. Each distribution is separately normalised and the $sin b$ has been randomised for quasars and galaxies (constant probability
per unit sky area).}
\label{fig:Data-Feature_Hists}
\end{figure*}

\begin{figure}[]
\centering
\includegraphics[]{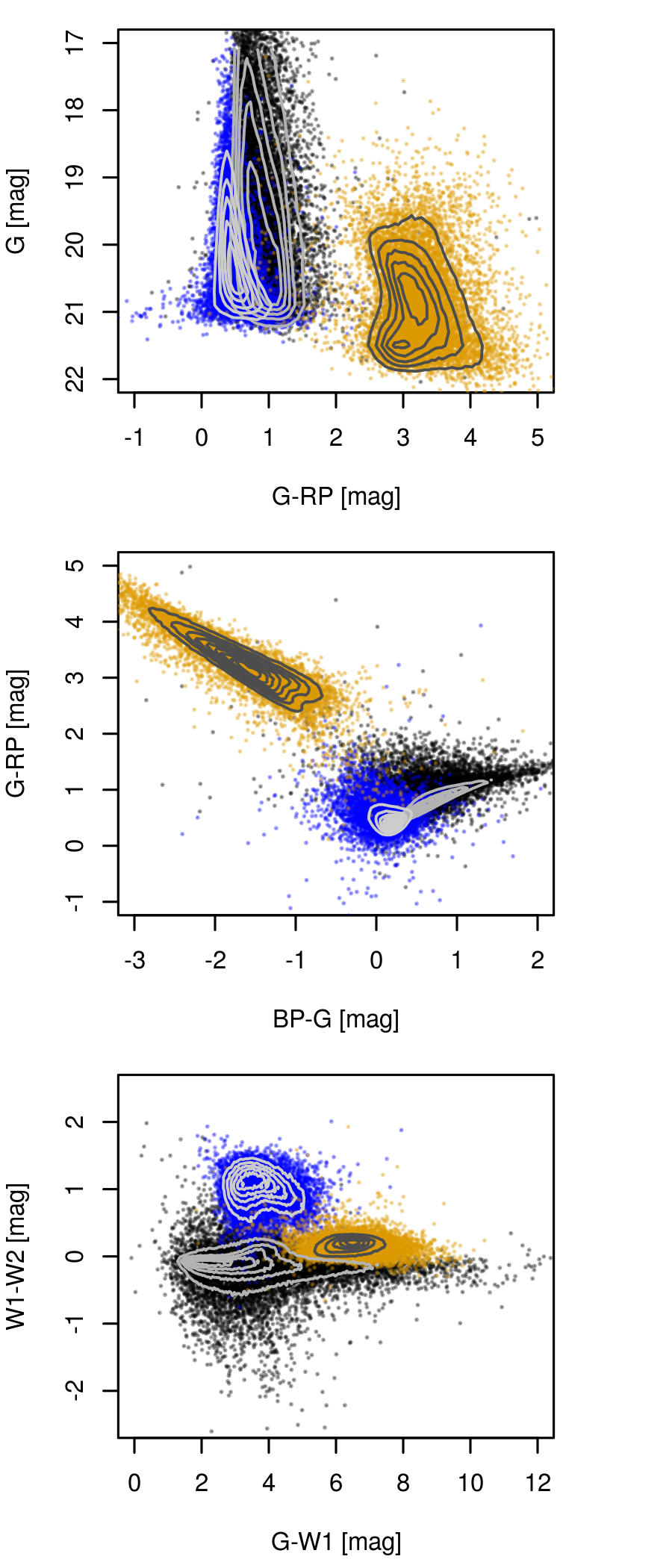}
\caption{Colour- magnitude diagram (\textit{top}) and two colour-colour (\textit{middle \& bottom}) diagrams highlighting the distribution of each class, with contours designating the density on a linear-scale for a random sample of 10,000 observations. The color black corresponds to stars, blue for quasars and gold for galaxies. Distinct aggregates can be identified for each class, although a significant interclass overlap still occurs.}
\label{fig:Data-col_col}
\end{figure}

\begin{itemize}

\item[-]
     Figure~\ref{fig:Data-Feature_Hists} shows the distribution of the broadband G magnitude in Gaia and the colours BP-G, G-RP, W1-W2 and G-W1. 
     Quasars have characteristic optical-IR colours. In the colour-colour and colour-magnitude space, quasars can be discerned from other stellar objects as well as from galaxies, in Fig.~\ref{fig:Data-col_col}. Additionally in Fig.~\ref{fig:Data-col_col}, we can see the clear distinction from the galaxy class.
     Due to the clear separation between the distinct classes seen in  Fig.~\ref{fig:Data-Feature_Hists} and Fig.~\ref{fig:Data-col_col}, we consider the colour information as one of the main discriminating features of the target classes.
 \item[-]  
    Galactic longitude and latitude $(l,b)$ can also be useful discriminants. Compared to stars, for which the distribution is concentrated towards the Galactic disk and the bulge, extragalactic objects are expected to be uniformly distributed across the entire sky (Copernican principle). However, such distribution is not observed due to the strong interstellar extinction in the disk of the Galaxy concealing extragalactic sources at low latitudes. Due to the SDSS sky coverage, the distribution of extragalactic objects in our train/test datasets follow a non-uniform distribution. We corrected our sample from this selection effect by randomising the latitude of these objects in our train and test datasets with values drawn from a uniform distribution in $sin b$. This approach may not be a perfect solution because, as just mentioned, we do not expect to see a large fraction of extragalactic objects at low latitudes. While this may help us find otherwise-difficult-to-detect extragalactic objects at low latitudes, it may also lead to a higher number of false positives. We accept this limitation. Galactic longitude is a problematic parameter because it wraps at $l = 0\degree = 360\degree$  and is not used as a model feature. However, we do use $l$ when computing our priors to account for the footprint of SDSS in comparison with Gaia (see Sect.~\ref{subsubsec:lat-magprior}).
 \item[-] 
    The relative variability in the G-band, which we call "relvarg" following the work by \cite{bailer-jones_quasar_2019}, is defined as the ratio of the standard deviation of the epoch photometry to its mean. Relvarg can be computed from the fields in GeDR3 as phot\_g\_n\_obs/phot\_g\_mean\_flux\_over\_error. Figure~\ref{fig:Data-Feature_Hists} shows a higher variability in quasars compared to stars. Galaxies also show large levels of variability, although in Gaia this effect is a spurious artefact due to galaxies being extended in their surface-brightness profile. At each epoch scan, Gaia will determine a slightly different photocentre possibly related to a different photometry. We can nonetheless exploit this behaviour to help distinguish galaxies.   

\item[-] 
    The astrometric unit weight error, UWE, is defined as the square root of the $\chi^2$ by the number of degrees of freedom of the astrometric solution. A larger UWE value correlates to a weak fit to the astrometric solution and generally an enhanced value for some galaxies. We do not use the re-normalised UWE (RUWE), which also removes dependencies on colour and magnitude, because RUWE is not defined when the parallax and proper motions are missing.

\end{itemize}



\section{Classification}
In our work, the goal of the classification task is to find an optimal mapping between a class label (i.e., star, quasar or galaxy) and a set of features characteristic of a given object. Several methods proposed in the literature have exploited supervised classification to determine the best mapping between input features and discrete classes. In our work, we seek a probabilistic classification, whereby a trained classifier generates a probability that an object belongs to a class, offering more flexibility in the determination of the final class prediction. Moreover, exploiting a Bayesian framework allows, on one hand, to define the posterior probability for an object to belong to a specific class and on the other hand to incorporate the use of a prior function highly informative on the target classes, such as the expected distribution of objects across the universe.

The following section introduces the terminology and the classification metrics used in our evaluation, the probabilistic models used to identify extragalactic sources, in addition to the prior functions we exploit to address the issue of class imbalance.

\subsection{Terminology and metrics}\label{Sec:Terminology}
In this section we define key terms and the metrics used to assess a classifier performance.
Classes may be defined as true and predicted. The true class refers to what has been defined in the training and testing datasets as the object's assumed class as defined by SDSS (for galaxies and quasars) and the stars by Gaia, and is therefore bound to have some inherent misclassification errors which will add noise to our classifier. The predicted class refers to the class that has been assigned using the probabilities outputted from our estimated classifier, which may be taking the maximum posterior probability or by considering a probability threshold. We define our predicted class as being the maximum posterior probability for a given source.
To compare the predicted and true classes, we construct a confusion matrix, where entry row $i$ and column $j$ of the matrix refer to the number of objects with the true class $i$ classified into predicted class $j$. The confusion matrix is of dimension $K \times K$, with $K^{2}$ numbers when classified using the maximum posterior probability. 

During training, we seek to minimise a loss function and monitor the performances of the model across all iterations using an evaluation metric. In multiclass classification, the standard loss function is the cross-entropy, defined in Eqn.~\ref{Equation:Cross entropy loss}, for which an ideal model would be able to correctly predict all objects (i.e., a cross entropy loss value equal to 0), in contrast with the opposite case of a larger value when the predictions diverge from the true class.

\begin{equation}
    \mathrm{Cross\ entropy\ loss}  = \frac{-1}{N}\sum_{n=1}^{N}\sum_{j=1}^{K}y_{nj}\log (p_{nj})
    \label{Equation:Cross entropy loss}
\end{equation}
where $N$ refers to the sample size, $K$ the number of classes, $y_{nj}$ the outcome equal to 1 for the true class and 0 otherwise, and $p_{nj}$ the probability that object $n$ belongs to the class $j$.

Classification performances are evaluated on a dataset unseen during the training phase, i.e.\ the test dataset.
Performances are evaluated through metrics such as the purity, the completeness and the F1-score. The purity, also known as precision (Eqn.~\ref{Equation:Purity}), refers to the number of true positives (TP) over the full count of objects in the target class. Purity can also be considered as a measure of contamination, (1 - purity), representing  the false positive rate (FP). The higher the purity the lower the contamination.
\begin{equation}
    \rm{Purity} = \frac{TP}{TP + FP}
    \label{Equation:Purity}
\end{equation}
The completeness, also known as the recall or sensitivity (Eqn.~\ref{Equation:Completeness}) refers to the number of true positives over the number of objects in the target class, i.e.\ the total sum of correct predictions (true positives) and true non-detections (false negatives). A perfect model has a purity and completeness both equal to 1.
\begin{equation}
    \rm{Completeness} = \frac{TP}{TP + FN}
    \label{Equation:Completeness}
\end{equation}

The F1-score is computed as the harmonic mean of a model's completeness and purity:
\begin{equation}
    \rm{F1} =2 \times \frac{Purity \times Completeness}{Purity + Completeness }
    \label{Equation:F1}
\end{equation}
We define the objective function during training as the cross-entropy (Eqn.~\ref{Equation:Cross entropy loss}), but we use the F1-score as the evaluation metric applied to the validation dataset for the statistical methods described in Sect.~\ref{Sec:Stat methods}. A perfect model has an F1-score of 1. 

We report the completeness and purities in the discussion of each of our classifiers, as these metrics are of most interest when considering the rare classes, quasar and galaxy, and because these objects are harder to classify in comparison to the large number of stars observed in Gaia.

\subsection{Statistical methods}\label{Sec:Stat methods}

Gaussian Mixture Models (GMMs) and Gradient Boosting methods have been shown to be effective in numerous classification tasks, such as the works by \cite{lee_application_2012,de_souza_probabilistic_2017} and \cite{moller_photometric_2016,chao_study_2019,golob_classifying_2021} respectively.
In the current section, we describe both methods as well as their known limitations when applied to our classification problem.

\subsubsection{Gaussian Mixture Models}
Gaussian Mixture Models \citep{fraley_model-based_2002}, GMMs, used in the work by \cite{bailer-jones_quasar_2019} for the supervised classification of extragalactic sources in Gaia, are defined in this work as our baseline classifier. In the training phase, the GMMs fit for each class of the training set the distribution of the data as a sum of $M$ Gaussians defined in a multi-dimensional feature space by maximum likelihood. In the prediction phase, for an unclassified object, the trained classifier computes a probability density function normalised for each class to provide posterior class probabilities, which nominally is equivalent to adopting an equal class prior. The final class prediction is obtained from the highest posterior probability across all classes.

GMMs are known to reach their limitations when dealing with overlapping classes and sparse data. To prevent such limitations, we introduced an adjustment to the likelihood by setting a fraction $n$ of the lowest values to zero, which sets the final densities computed by the GMMs to zero. By forcing the Gaussian distributions to truncate to zero, sources at the boundaries of a class distribution (the potential overlap between classes) should be directly assigned to the most prevalent class (for our purpose the star class), resulting in an increase in the purity and completeness of the rarer classes. We considered four different values for the threshold value $n$ of \{1,5,20,50\} applied to all models trained on the different input configurations (i.e., with and without infrared features). We found that the purity in the quasar and the galaxy classes marginally improves (an increase of 0.02) when using $n=50$ for the model trained without the infrared features compared to a standard GMM. However, this correction does not induce any improvement for the models trained on the dataset including infrared features. Furthermore, we find that the GMMs subject to the likelihood trimming method perform better in contrast with the standard GMM classifier, but attain lower performances compared to the boosted decision tree methods. We therefore do not consider the correction via likelihood trimming further.

\subsubsection{Gradient boosting methods}
Gradient boosting is a popular and powerful ensemble technique within supervised machine learning, where the ensemble technique refers to building a model from a collection of weaker learners. There are additional ensemble methods such as bagging that splits the dataset into $N$ subsets with replacement, builds a model on each subset in parallel and finally combines their individual predictions to compute the final class. 
Bagging is the basis of the Random Forest method \citep{breiman_random_2001}.
By contrast, gradient boosting builds a model by sequentially fitting the weak learners in order to correct the residual errors at each iteration. The algorithm re-weights the data towards the most difficult cases at each training step, such that subsequent learners prioritise solving them. Typically, the learners used in gradient boosting methods are decision trees, and the method is known as Gradient Boosting Decision Trees (GBDT) \citep{friedman_greedy_2001}.

The eXtreme Gradient Boosting method, namely XGBoost, refers to a boosting algorithm presented in \cite{chen_xgboost_2016} refers to one of the fastest implementation of GBDT. In particular, where XGBoost improves upon GBDT lies in the inclusion of the second derivative of the loss function, which provides complementary information on the direction of gradients essential to solve the optimisation problem. Furthermore, the XGBoost method includes L1 and L2 regularisation used to prevent the model from overfitting.

A second gradient boosting method, used in our work, is Categorical Boosting (CatBoost) \cite{dorogush_catboost_2017}. The key properties of CatBoost compared to XGBoost are; balanced (symmetric) trees and ordered boosting. Balanced trees, by definition, are built such that, at every step, the trees are split using the same feature criterion. By using a balanced tree architecture, CatBoost runs more efficiently and controls for overfitting as the balanced tree serves as regularisation. In general, classic boosting methods are prone to overfitting and CatBoost circumvents this limitation via ordered boosting, which refers to the process of training a model on a subset of the data and computing the residuals on a different subset. 

In the following, we train two classifiers, using XGBoost and CatBoost, with the similar set of input features used to train the GMM in order to assess the classification performances of all classifiers. We select the optimal hyperparameters by performing a 5-fold grid search cross-validation that minimises the cross entropy loss function for finding the best hyperparameters. We then maximise the evaluation metric, the F1 score, when fitting the model with the best hyperparameters on the validation set. The hyperparameters we choose to optimise are $max\_depth$ or $depth$ (in the case of CatBoost) , $learning\_rate$ and $n\_estimators$, with the remaining set at their default values. $max\_depth$ represents the maximum number of nodes allowed on a tree and is used to control for over-fitting as a higher depth will make the model more complex and representative of the training dataset and thus more likely to be overfitted. The $max\_depth$ parameter ranges from 0 to infinity and we consider the values of 3, 6, 8 and 10. The $learning\_rate$ is the step size the model takes at each iteration to reach the minimum of the loss function, takes a value between 0 and 1, and is used to control for overfitting by modifying the weights of new trees added to the model. We consider the values 0.01, 0.03, 0.05 and 0.1. The last hyperparameter we consider tuning is the number of trees, specified by $n\_estimators$. There is often a point of diminishing returns once we have a large number of trees, when each subsequent tree hardly reduces the loss function. We considered the values of 100, 500 and 1000 for the $n\_estimators$ in our testing.
The optimal values for the hyperparameters obtained from our grid search are a $max\_depth$  of 8, a $learning\_rate$ of 0.1  and a total number of trees $n\_estimators$ of 100 for XGBoost, and for CatBoost a $depth$ of 6, a $learning\_rate$  of 0.03 and a total number of trees $n\_estimators$ of 1000.

\subsection{Prior}\label{Subsec:The prior}
The class imbalance problem, when the class distributions are highly skewed and we are interested in the less frequent class, is not unique to classification within astronomy. The problem is often encountered in various areas such as credit fraud detection where fraud is considerably less frequent than regular transactions. Multiple classification algorithms in this context attain low predictive accuracy for the rare class. Several data augmentation methods have been developed to address the imbalance problem from oversampling the rare classes, undersampling the most prevalent class and generating synthetic observations using techniques such as SMOTE \citep{chawla_smote_2002}. In this work, we attempt to correct for the class imbalance by applying a model correction exploiting prior knowledge that can be physically attributed, as introduced in~\cite{bailer-jones_quasar_2019}. \cite{lake_exploration_2022} offer a similar approach which likewise proposes replacing the implicit prior of the classifier with one representative of the target population. 
 The model correction is applied in two phases of the modelling process. First by adjusting the posterior probabilities by the class prior, as described in Eqn.~\ref{eq:Post_probs adjust}, and second via the modification of the confusion matrix by an adjustment factor, $ \lambda_{k}$, shown in Eqn.~\ref{eq:adjustment factor}. The approach is thoroughly explained in section 3.4 of ~\cite{bailer-jones_quasar_2019}, however, we summarise key points in the following section for convenience. 
 
 First, the prior adjustment is done by re-weighting the posterior probabilities using a prior distribution to reflect the expected real class distribution. 
 \begin{equation}\label{eq:Post_probs adjust}
     P(C_{k}|x,\Theta) = \frac{1}{Z}\pi_{k}P(x,|C_{k})
 \end{equation}
where\,$\Theta$\, refers to any prior information, $Z = \sum_{k} \pi_{k}P(x,|C_{k})$ and  $\pi_{k}$ is the class prior for class $k$.

Second, when applying the model to a test dataset, the confusion matrix is modified using the adjustment factor in Eqn.~\ref{eq:adjustment factor}. This approach ensures that the results would reflect the expected (prior) distribution of all classes, in particular the larger number of potential star contaminants to the quasar and galaxy classes. This step is necessary because the actual test dataset generally does not portray the class distribution expected in reality; in particular it will tend to have too few stellar contaminants. 
The factor $ \lambda_{k}$ scales the actual number of objects in each row to the number of objects expected within a dataset. Given the definition of the adjustment factor, the correction only affects the purity and not the completeness estimated from the confusion matrix.
\begin{equation}\label{eq:adjustment factor}
    \lambda_{k} = 
        \frac {\pi_{k}}{\alpha_{k}}
            \left ( \sum_{k`} (\frac{\pi_{k`}}{\alpha_{k`}}) 
            \right )^{-1}
\end{equation}
where $\alpha_{k}$ is the class fraction within a dataset.\\

In this work, we describe three different priors and apply two of them, first, the global prior reflecting a general class distribution, and secondly, a joint prior dependent upon latitude and magnitude, and lastly, a mixed prior that combines the two aforementioned priors.

\subsubsection{The global prior}
The global prior of $\{\pi_{star}^{GP},\pi_{quasar}^{GP},\pi_{galaxy}^{GP}\}$, introduced in the work by \cite{bailer-jones_quasar_2019}, outlines the scarcity of quasar and galaxy objects compared to stars across the sky. The prior sets the probability of observing a quasar to 1/1000 and to 1/5000 for galaxies from a sample of extragalactic sources with parallaxes and proper motion measurements.
However, as discussed in Sect.~\ref{subsubsec:galaxies}, the majority of galaxies observed in Gaia lack reported parallaxes and proper motions. To define our global prior, we count the number of sources across each class in the Stripe82 region from SDSS DR16, and extrapolate the distribution across the entire sky. 
The SDSS region Stripe 82 is chosen given the large sample of spectroscopic observations available for the majority of sources, thus providing a more complete count of identified targets. We find twice the number of galaxies compared to quasars and based on this we define our global prior as 1/1000 for quasars and re-adjusted to 1/500 for galaxies.

\subsubsection{The joint latitude and magnitude prior}\label{subsubsec:lat-magprior}
Extragalactic sources are expected to have an intrinsic uniform distribution across the sky, but will not be observed due to dust extinction in the disk.
We would expect at low latitudes, closer to the galactic plane, a higher number density of stars in comparison to galaxies and quasars. As the (absolute) latitude increases, the number density of galaxies and quasars to stars increases. This information can be used to generate a latitude-based prior, derived from densities at different latitudes.

We can also construct a prior based on apparent magnitude as we would expect the number of quasars and galaxies to increase towards the fainter brightness end. The G-band magnitude distribution in Fig.~\ref{fig:Data-Feature_Hists}, supports this expectation. Exploiting such characteristics in the latitude and the G magnitude distributions, we have the functionality to represent what we consider to be true variations in latitude and magnitude as a prior to improve the performance of our classifier over a 2-dimensional (2D) latitude and magnitude space.

To construct the joint class prior, we choose the overlapped region $50^\circ<= l <= 200^\circ$ in Gaia and SDSS DR16, to ensure that we are counting sources over the same area of the sky. We assume that SDSS DR16 includes all galaxies and quasars in this region, and that Gaia includes all stars within. Here the denomination "all" refers to a randomly generated application data set (i.e., our randomly-selected 50 million GDR3 sources). 
Using the compiled list of sources, we now further define bins in both $sin b$ and $G-mag$, count the number of stars, quasars, and galaxies and finally normalise in order to compute frequencies.
The distributions for the different class priors can be seen in Fig.~\ref{fig:Data-priors}. The top panel shows a large number of stars within the plane and a lower density of stars at higher latitudes and towards lower magnitudes. The middle panel reports the distribution observed for the quasars, for which the lowest density is identified within the lowest latitude bin, and a majority of quasar sources at  $G=18$\,mag and higher latitudes. For galaxy sources, the bottom-panel reports the majority of sources at the highest magnitudes and uniformly distributed across latitudes excluding the lowest latitude regions.

\subsubsection{The mixed prior}\label{mixedprior}
\begin{figure}[]
\centering
\includegraphics[scale=0.7]{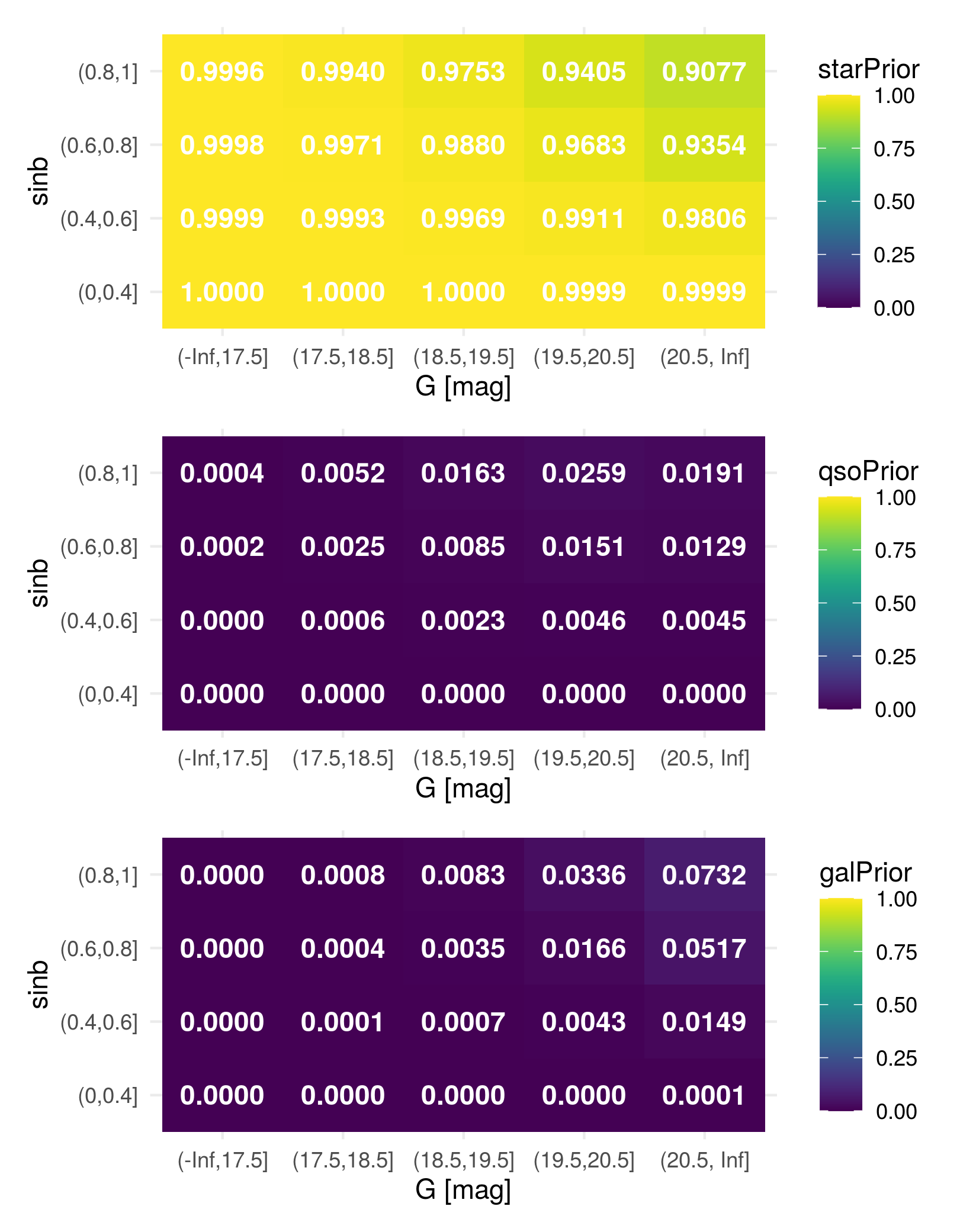}
\caption{Heat map of the mixed prior distribution. In this representation, the number of stars at lower latitudes exceeds the number of observed quasars and galaxies. Whereas, at higher latitudes and fainter magnitudes, the number of quasars and galaxies compared to the number of stars increases. Values of "0.0000" are not [necessarily] exactly zero, but below the numerical precision shown. }
\label{fig:Data-mixedpriors}
\end{figure}

The mixed prior refers to the latitude and magnitude dependent prior that accounts for the overall sky distribution of classes represented by the global prior.
We define the mixed prior as follows.
\begin{enumerate}
    \item $g_{S},g_{Q},g_{G} \simeq (1,\frac{1}{1000},\frac{1}{500})$, the (unnormalised) target global prior.
    \item $F_{S},F_{Q},F_{G}$ are the measured fraction of sources by star, quasar, and galaxy class in the overlap of SDSS and Gaia over the region $50^\circ<= l <= 200^\circ$, over all latitudes and magnitudes.
    \item In a specific latitude and magnitude bin the number of sources of each class are counted to be $n_{S},n_{Q},n_{G}$. 
    \item The number of sources we should have in each latitude and magnitude bin according to our target prior are therefore \\ $n_{S}^{'}=n_{S}\frac{g_{S}}{F_{S}} \\ 
    n_{Q}^{'}=n_{Q}\frac{g_{Q}}{F_{Q}} \\
    n_{G}^{'}=n_{G}\frac{g_{G}}{F_{G}}$ . 
    \item Normalizing these across the classes gives the target prior for each latitude and magnitude bin:\\
    $n_{S}^{''}=\frac{n_{S}^{'}}{n_{S}^{'} +  n_{Q}^{'} +  n_{G}^{'}} \\ 
    n_{Q}^{''}=\frac{n_{Q}^{'}}{n_{S}^{'} +  n_{Q}^{'} +  n_{G}^{'}} \\ 
    n_{G}^{''}=\frac{n_{G}^{'}}{n_{S}^{'} +  n_{Q}^{'} +  n_{G}^{'}}$ \\ 
\end{enumerate}
The distribution of this prior across latitude and magnitude is shown in Fig.~\ref{fig:Data-mixedpriors}. We see the dominance of stars in the lower latitudes and a gradual increase in prevalence of quasars and galaxies at higher latitudes and fainter magnitudes. As the prior is discontinuous in magnitudes $G$ and latitudes $b$, we expect  discontinuities in the classification probabilities and counts.


\begin{table*}[!ht]
\centering
\caption{Classification performances obtained for different balanced classifiers using different algorithms and input features. Completeness and purity are shown for each class. From our tests, the best performing model is the XGBoost algorithm trained on the Gaia\_f features supplemented with the infrared CatWise2020 colours (\textit{Feature Set 4})}
\begin{tabular}{cllllllll}
 & & \multicolumn{2}{c}{Completeness} & \multicolumn{3}{c}{Purity} & \\
\toprule
& Features & Star & Quasar & Galaxy & Star & Quasar & Galaxy \\ 
\hline
\multirow{4}{*}{\rotatebox{90}{GMM}}
    & Gaia\_f     
        &  0.9330 &	0.9580 &	0.9886 
        &	0.9532 &	0.9405 &	0.9860 \\ 
    & Gaia\_f + W1-W2 
        & 0.9714 &	0.9850 &	0.9906 
        &	0.9810 &	0.9784 &	0.9875 \\ 
    & Gaia\_f + W2 + G-W1 
        & 0.9766 &	0.9871 &	0.9919 
        & 0.9853  &	0.9837 &	0.9866 \\ 
    & Gaia\_f + W1-W2 + G-W1 
        &  0.9778 &	0.9859  &	0.9919 
        &	0.9840 &	0.9846 &	0.9869 \\ \bottomrule
\multirow{4}{*}{\rotatebox{90}{XGBoost}} 
    & Gaia\_f     
        &  0.9418 &	0.9623 &	0.9922 
        &	0.9603 &	0.9489 &	0.9871 \\ 
    & Gaia\_f + W1-W2 
        & 0.9728 &	0.9878 &	0.9933 
        &	0.9857 &	0.9798 &	0.9885 \\ 
    & Gaia\_f + W2 + G-W1 
        & 0.9793 &	0.9896 &	0.9932 
        &	0.9878 &	0.9859 &	0.9884 \\ 
    & Gaia\_f + W1-W2 + G-W1 
        & 0.9793 &	0.9908 &	0.9936 
        &	0.9891 &	0.9857 &	0.9889  \\ 
\bottomrule
\multirow{4}{*}{\rotatebox{90}{CatBoost}} 
    & Gaia\_f     
        & 0.9411 &	0.9619 &	0.9919 
        &	0.9593 &	0.9484 &	0.9872 \\ 
    & Gaia\_f + W1-W2 
        & 0.9720 &	0.9876 &	0.9930 
        &	0.9854 &	0.9787 &	0.9885  \\
    & Gaia\_f + W2 + G-W1 
        & 0.9785 & 0.9883 &	0.9927 
        &	0.9862 &	0.9847 &	0.9886 \\ 
    & Gaia\_f + W1-W2 + G-W1 
        & 0.9786 &	0.9905 &	0.9934  
        &	0.9886 &	0.9850 &	 0.9888  \\ 
\bottomrule
\end{tabular}
\label{Table1:Balanced Classifier}
\end{table*}

\begin{table*}[!ht]
\centering
\caption{Classification performances adjusted by the global prior and adjustment factor for different balanced classifier models using the XGBoost algorithm and different input features.}
\begin{tabular}{cllllllll}
 & & \multicolumn{2}{c}{Completeness} & \multicolumn{3}{c}{Purity} & \\
\toprule
& Features & Star & Quasar & Galaxy & Star & Quasar & Galaxy \\  
\hline
\multirow{4}{*}{\rotatebox{90}{XGBoost}} 
    & 1: Gaia\_f     
        &  0.9992 &	0.0844 &	0.3625 
        &	0.9978 &	0.5295 &	0.4915 \\ 
    & 2: Gaia\_f + W1-W2 
        & 0.9987 &	0.2653 &	0.4261 
        &	0.9981 &	0.4209 &	0.4768 \\
    & 3: Gaia\_f + W2 + G-W1 
        & 0.9986 &	0.3402 &	0.4464 
        &	0.9982 &	0.4197 &	0.4924 \\ 
    & 4: Gaia\_f + W1-W2 + G-W1 
        & 0.9986 &	0.3289 &	0.4650 
        &	0.9983 &	0.3944 &	0.5054  \\ 

\bottomrule
\end{tabular}
\label{Table1:Balanced Classifier - Adjusted}
\end{table*}

\begin{table*}[]
\centering
\caption{Classifier using XGBoost with two feature set configurations as applied to the imbalanced test dataset. Classification performances of the model trained on an imbalanced dataset shows lower completeness but higher purities compared to the balanced classifier. Classification performances increase when the infrared data is incorporated as input features.}
\begin{tabular}{llllllll}
 &  \multicolumn{2}{c}{Completeness} & \multicolumn{3}{c}{Purity} & \\
\toprule
Features & Star & Quasar & Galaxy & Star & Quasar & Galaxy \\
\hline
Gaia\_f  & 0.9714 &	0.9040  &	0.9914 & 0.9766  & 0.9091 &	0.9759 \\ 
Gaia\_f + W1-W2 + G-W1 & 0.9858    &	0.9799 &	0.9922 &	0.9937 &    0.9705  &	0.9784 \\ 
\bottomrule
\end{tabular}
\label{Table4:XGBoost-Maximum Source Classifier}
\end{table*}

\begin{table*}
\small
\centering
\caption{Confusion matrix on the test set predictions using an XGBoost classifier trained on \textit{Feature Set 4}. The right half of the table has been modified by the adjustment factor.}
\begin{tabular}{cllll|lll}
& & \multicolumn{1}{c}{Predicted} \\
\toprule
 & & Star & Quasar & Galaxy & Star & Quasar & Galaxy \\ \midrule
\multirow{3}{*}{\rotatebox{90}{Actual}}& Star & 887262 & 5127 & 7611   & 199119.6 & 100.6979 & 181.4556  \\ 
                                       & Quasar & 3529 & 195980 & 491 & 133.8076 & 65.5892 & 0.0050  \\ 
                                       & Galaxy & 2076 & 820 & 367104 & 213.3360 & 0.0260 & 185.4417  \\ \bottomrule 
\end{tabular}
\label{cm:og+w1-w2}
\end{table*}

\section{Results of different models and feature combinations on the test set}
Section~\ref{Subsec:Balanced results} presents the results of classification obtained with four different feature combinations using the GMM, XGBoost and CatBoost methods applied to the balanced data set (for training and testing). We identify the best feature combination and method for the classification of extragalactic sources.
Section~\ref{Subsec:Max Source Classifier} shows the results of the chosen model and features combination fitted and assessed on the larger imbalanced training and test datasets (respectively). The effect of applying the priors to the model probabilities is discussed in Sect.~\ref{Sec:PriorAdjusted}. The selected classifier is applied to our application datasets in Sect.~\ref{Sec:Application}.

Our tests were run on an Ubuntu server, with 344GB of RAM and an Intel Xeon CPU E5-2695 v3 at 2.30 GHZ with 30 threads.

\subsection{Classifier trained on a balanced set}\label{Subsec:Balanced results}
By considering a balanced class distribution across the training and test datasets with 200\,000 sources in each class, as defined in Sect.~\ref{Subsec:Training and test}, the intrinsic performances of each method applied to different input feature combinations are higher compared to the results obtained when we subsequently apply the appropriate prior and allow for a higher level of contamination from stellar objects.

Table~\ref{Table1:Balanced Classifier} reports the different methods GMM, XGBoost, and CatBoost, where each model is tested using four combinations of input features:

\begin{itemize}
    \item[] \textit{Feature Set 1}: Gaia\_f 
    \item[] \textit{Feature Set 2}: Gaia\_f + W1-W2
    \item[] \textit{Feature Set 3}: Gaia\_f + W2 + G-W1
    \item[] \textit{Feature Set 4}: Gaia\_f + W1-W2 + G-W1
\end{itemize}

When performing model fitting, we search for the best input configuration with the highest purity and completeness in the quasar and galaxy classes. We add the colour difference of W1-W2 and G-W1 for long colour wavelength span to the original Gaia features. We do not consider colours such as BP-W2, as W2 is less sensitive than W1 and G has a higher signal-noise ratio than BP.
From Table~\ref{Table1:Balanced Classifier}, we can see that across all feature combinations, the GMM has a lower classification performance for the two extragalactic classes compared to the gradient boosted methods. Moreover, the addition of infrared derived features increases the purity and completeness for the quasar and galaxy classes. In Table~\ref{Table1:Balanced Classifier - Adjusted}, we apply a global prior and the adjustment factor, and report the purity and completeness for each class. We find that Feature sets 3 and 4 give comparable performances, and are better than sets 1 and 2. Given Table~\ref{Table1:Balanced Classifier} and Table~\ref{Table1:Balanced Classifier - Adjusted} we choose an XGBoost model with Gaia features and the W1-W2 and G-W1 infrared colours (\textit{Feature Set 4}).

\subsection{Classifier trained on an imbalanced set}\label{Subsec:Max Source Classifier}

Using the statistical model and features identified in Sect.~\ref{Subsec:Balanced results}, we now train a classifier using all available sources. This enables the design of a classifier that is more representative of the true class distribution, as discussed in Sect.~\ref{Subsec:Training and test}. Introduced in Sect.~\ref{Subsec:The prior}, the global prior is set to (1, 1/1000, 1/500) respectively for star, quasar, and galaxy targets which differs from the class fractions in the training and test sets. Given the available data, it would be infeasible to use this prior and have a representative number of objects in the extragalactic classes. Furthermore, the intrinsic prior of a model is not necessarily equal to the class fractions in the data initially trained on. The discussion of applying the adjustment to the posterior probabilities is discussed in Sect.~\ref{Sec:PriorAdjusted}.

The results of our classifier trained using \textit{Feature Set 4} are reported in Table~\ref{Table4:XGBoost-Maximum Source Classifier}. We compare the final model with an XGBoost model trained exclusively on \textit{Feature Set 1} and find a significant improvement in the completeness and purity for the quasar class, from 0.9040 to 0.9799 in the completeness, and from 0.9091 to 0.9705 in the purity. However, for the galaxy class, only an insignificant improvement is seen in the classification metrics, from 0.9914 to 0.9922 in the completeness and from 0.9759 to 0.9784 in the purity. Compared to the balanced classifier in Sect.~\ref{Subsec:Balanced results}, the current classifier exploits a larger dataset, thus the decrease in the classification performances is to be expected particularly in the purity due to the fact that the larger dataset likely has more contaminants. 

For the remainder of this work, we retain the classifier trained on the imbalanced dataset using \textit{Feature Set 4} to assess the use of different priors applied to the models and apply the classifier to the application sets in Sect.~\ref{Sec:Application}.

\subsection{Classifier adjusted using the priors}\label{Sec:PriorAdjusted}

We now consider the effect of applying different prior probability distributions to the posterior probabilities estimated by the classifier in Sect.~\ref{Subsec:Max Source Classifier}. In the figures discussed in this section, the left-panels represent results obtained for the \textit{Feature Set 1} model, whereas, the right-panels report the results associated to the \textit{Feature Set 4}, both with XGboost.

The results using the global prior are reported in Table~\ref{Table:Results-GlobalPrior}. 
The top half of the table ("Adj") shows results for a realistic level of stellar contamination by using the adjustment factor $\lambda_{k}$ in Eqn~\ref{eq:adjustment factor}; the bottom half shows raw unadjusted results, i.e.\ with the lower level of contamination seen in the test set ("Unadj"). Using the global prior gives a lower completeness overall in comparison to the results obtained before applying the prior in Table~\ref{Table4:XGBoost-Maximum Source Classifier}. On average, similar results are observed in the purity for the unadjusted case. Applying the adjustment factor results in lower purities across both the quasar and galaxy class, however, the addition of infrared colour information clearly results in a better performing classifier.

\begin{figure*}[]
\centering
\includegraphics[width=\textwidth]{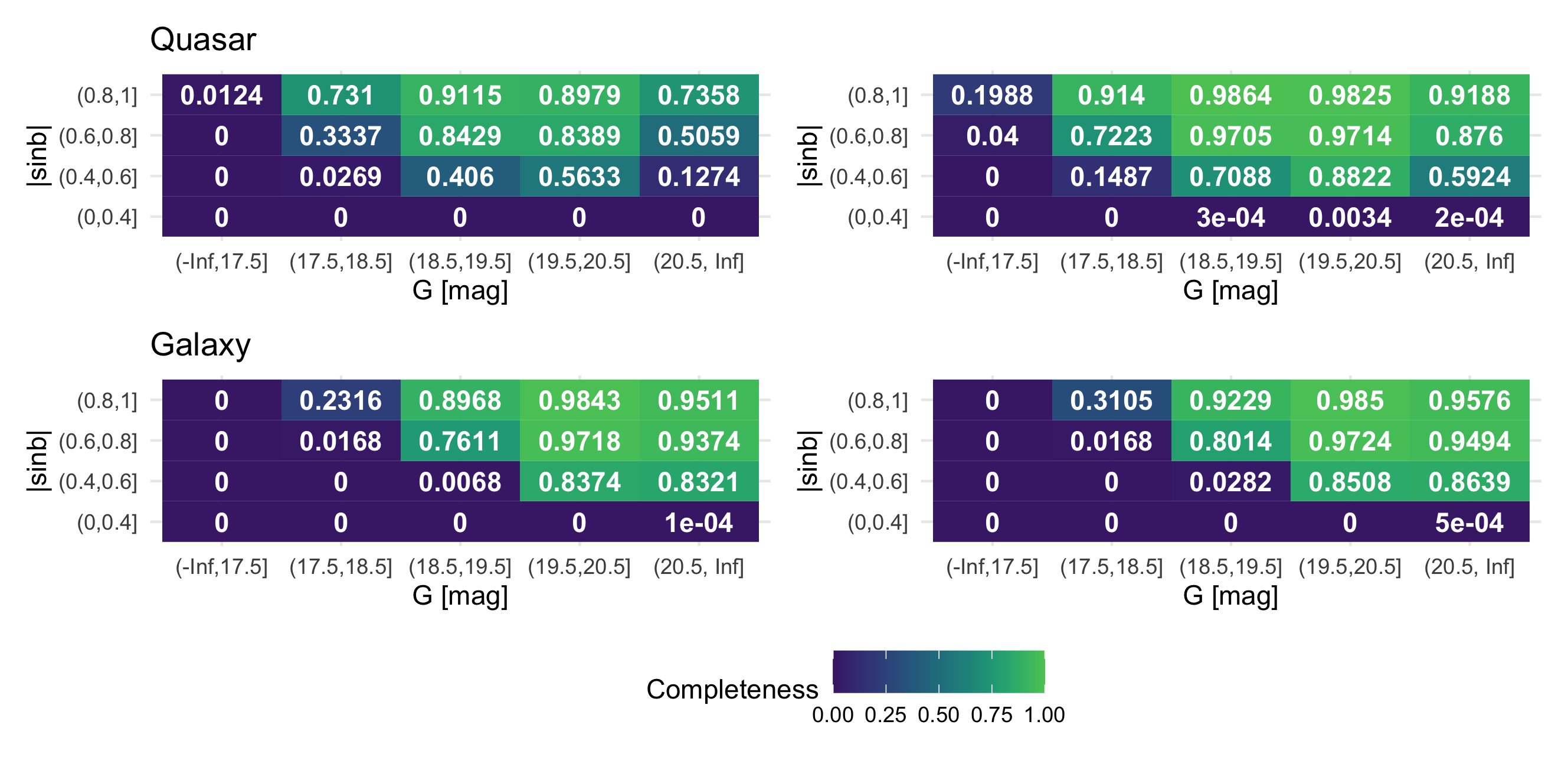}
\caption{Imbalanced Classifier Mixed Prior: Completeness evaluated for the three target classes in the test set from predictions obtained by the best performing models, i.e. XGBoost, trained on the \textit{Feature Set 1} (left panel) and the \textit{Feature Set 4} (right panel).
}
\label{fig:Results-3class_completeness_mixedprior}
\end{figure*}

\begin{figure*}[]
\centering
\includegraphics[width=\textwidth]{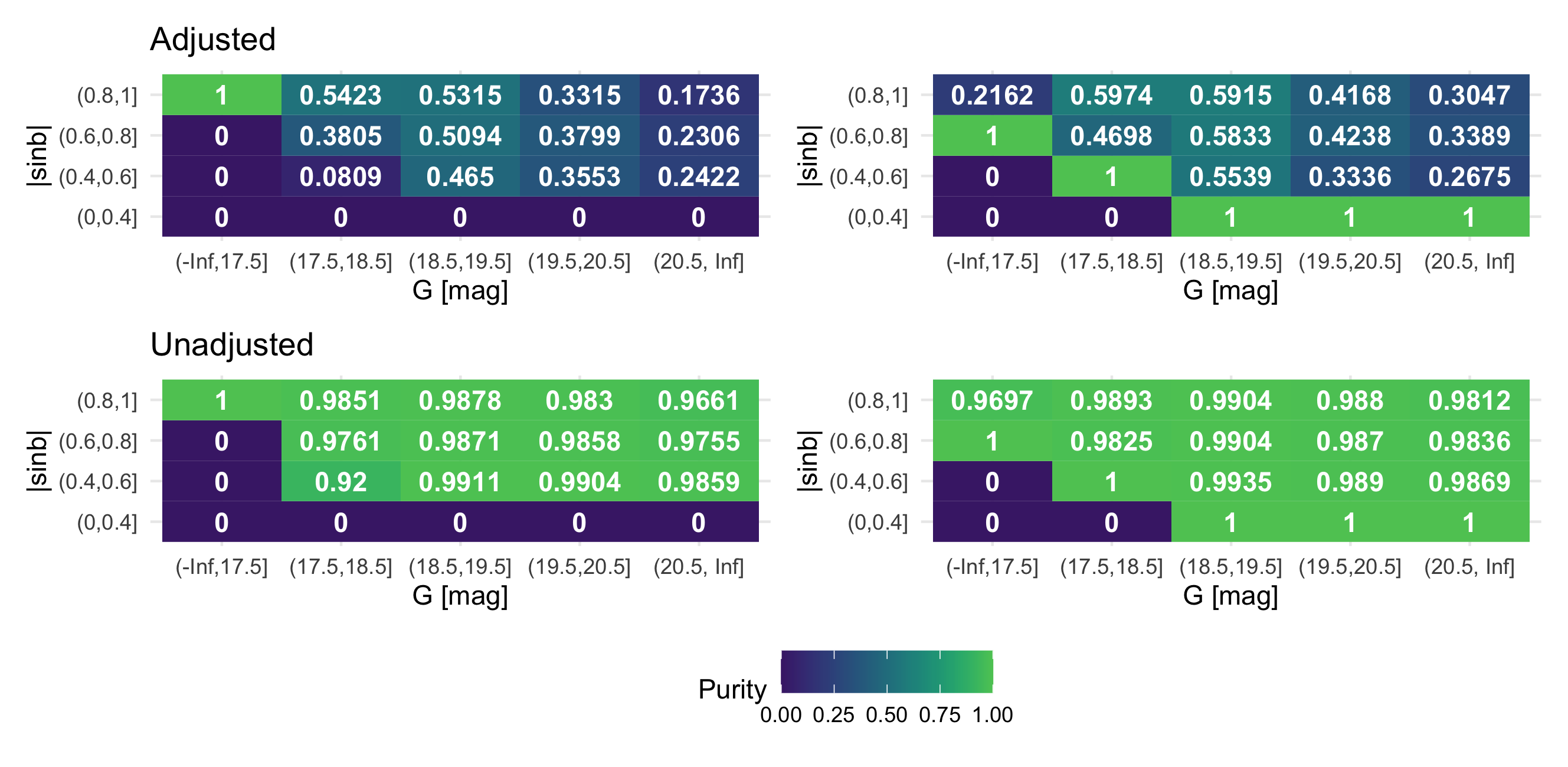}
\caption{Imbalanced Classifier Mixed Prior: Purity evaluated for the quasar class in the test set from predictions obtained by the best performing models, i.e. XGBoost, trained on the \textit{Feature Set 1} (left panel) and the \textit{Feature Set 4} (right panel). 
Top-panels show the classification performances modified by the adjustment factor. The near unit purity at low latitudes in the right panels is not meaningful as there a very few objects as shown in Fig.~\ref{fig:Data-mixedpriors}}
\label{fig:Results-3class_quasar_purity_mixedprior}
\end{figure*}

\begin{figure*}[]
\centering
\includegraphics[width=\textwidth]{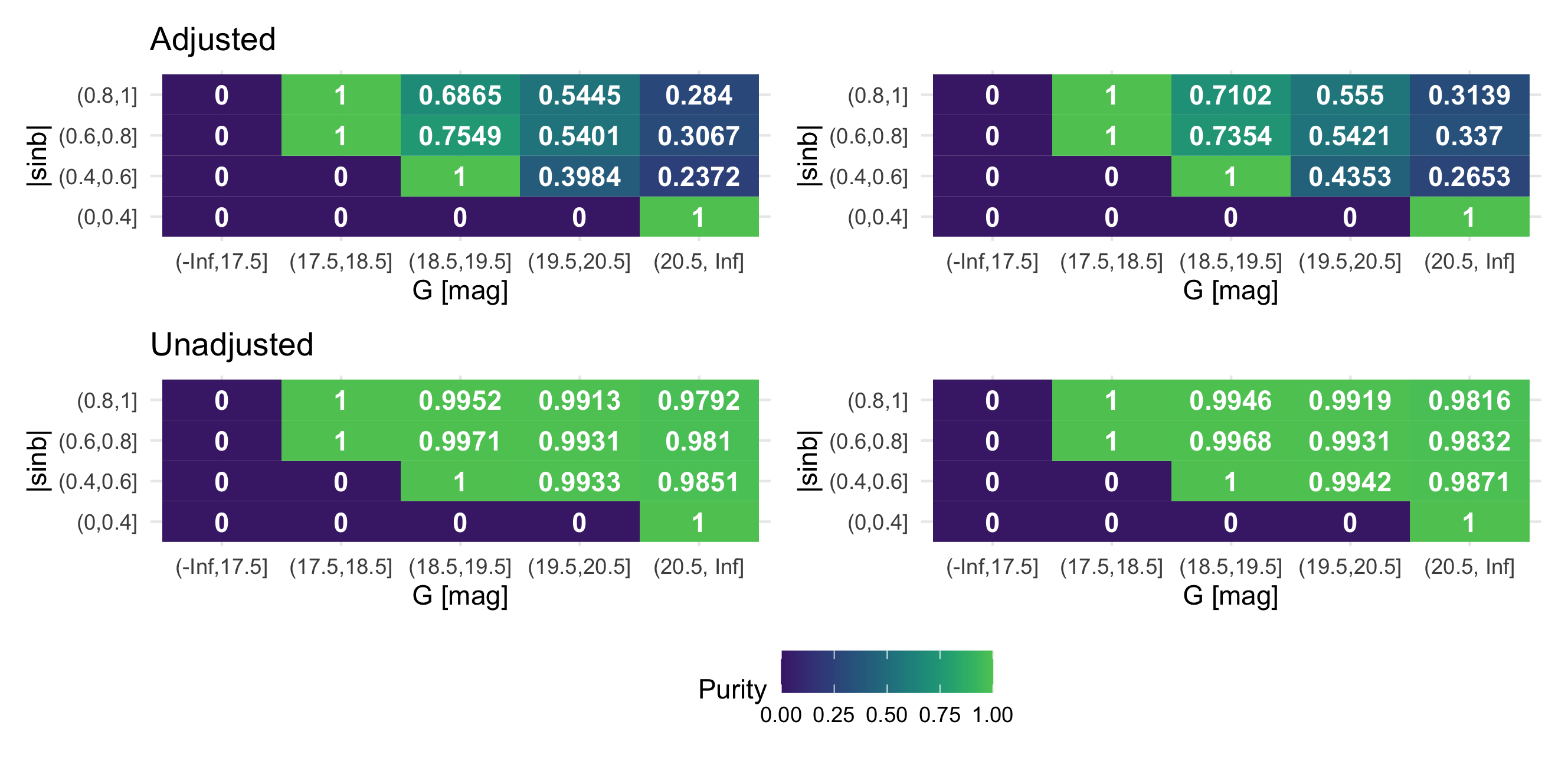}
\caption{Similar to Fig.~\ref{fig:Results-3class_quasar_purity_mixedprior} for the galaxy target class in the test set.}
\label{fig:Results-3class_galaxy_purity_mixedprior}
\end{figure*}

Having assessed the impact of the global prior on the final classification. We now consider a more tuned prior, namely the "mixed" prior introduced in Sect.~\ref{mixedprior}, and assess the performance as a function of latitude and magnitude, while also applying the adjustment of the confusion matrix in order to incorporate the expected class fractions at each latitude and magnitude into the performance metrics. 
In Fig.~\ref{fig:Results-3class_completeness_mixedprior}, the completeness for the quasar class improves with higher latitudes and most significantly when we add infrared colour information as an input feature. However adding infrared data and moving to higher latitudes marginally improves the completeness in the galaxy class.  As an illustration, there is an 18\% increase in completeness for very faint quasars at high latitudes (top-right bin) and only a 0.7\% increase in completeness for galaxies in the equivalent bin when adding infrared data. 
The purities for the quasar and galaxy classes are shown in Fig.~\ref{fig:Results-3class_quasar_purity_mixedprior} \& Fig.~\ref{fig:Results-3class_galaxy_purity_mixedprior} respectively. We would like to point out that the exact values of 1 and 0 are due to a rounding precision. The effect of the adjustment factor is reported in the top-panels. For the quasar class, we observe a significant improvement in purity when adding the infrared colours, and as a function of latitude.
For the galaxy class, the addition of infrared colours has only a marginal improvement on the purities as a function of latitude and magnitude. The application of the adjustment factor induces an expected decrease in the purity for both the quasar and galaxy target classes.

\begin{table*}[!ht]
\centering
\caption{Imbalanced Classifier Global Prior: Completeness and purity using the global prior as applied to the test dataset using the imbalanced classifier. "Adj" is defined as adjusted using the adjustment factor, $\lambda_{k}$,in Eqn~\ref{eq:adjustment factor} and "Unadj" without.}
\begin{tabular}{cllllllll}
 & &  \multicolumn{2}{c}{Completeness} & \multicolumn{3}{c}{Purity} & & \\
\toprule
 & Features & Star & Quasar & Galaxy & Star & Quasar & Galaxy   \\ 
\hline   
\multirow{2}{*}{\rotatebox{90}{Adj}} & Gaia\_f  &     0.9995 &  0.0897  & 0.3054  & 0.9977 & 0.4621  &   0.5958  \\ 
& Gaia\_f + W1-W2 + G-W1 &     0.9993  &   0.2131  &  0.3790 &  0.9980 & 0.5694 &  0.6036 \\ 
\bottomrule 
\multirow{2}{*}{\rotatebox{90}{Unadj}}& Gaia\_f  & 0.9995  &    0.0897   &      0.3054  & 0.6721  & 0.9946  &  0.9967 \\ 
& Gaia\_f + W1-W2 + G-W1 &     0.9993  &      0.2131  & 0.3790  & 0.6991  & 0.9962 &    0.9968  \\ 
\bottomrule
\end{tabular}
\label{Table:Results-GlobalPrior}
\end{table*}


\section{Results of the best performing model and feature combination on the application sets}\label{Sec:Application}

To evaluate how our selected classifier performs and what distribution of the predicted classes is obtained on datasets with representative distributions, we apply the classifier to three datasets selected from the 1.8 billion sources observed in Gaia at the intersection between GDR3 and the CatWISE2020 catalogue. Our first application aims to predict the classes for a randomly-selected subset of 50 million sources, without prior information on the target classes nor their distribution. This application set, however, has the distribution that our global and mixed priors are designed for.
The second dataset is constructed from the GDR3 quasar and galaxy candidate tables defined in ~\citep{gaia_collaboration_gaia_2022}, which are quoted as having purities of 0.52 and 0.69 respectively.
The third data set is the purer subsample of the candidates tables defined in ~\citep{gaia_collaboration_gaia_2022}, which are quoted as having purities of 0.95 for the quasar class and 0.94 for the galaxy class.
In addition to assessing the accuracy of our classifier, we wish to identify whether adding infrared colours to Gaia data improves the reliability of these candidate tables, despite having removed parallax and proper motion as features.

\begin{figure}[]
\centering
\includegraphics[scale=0.7]{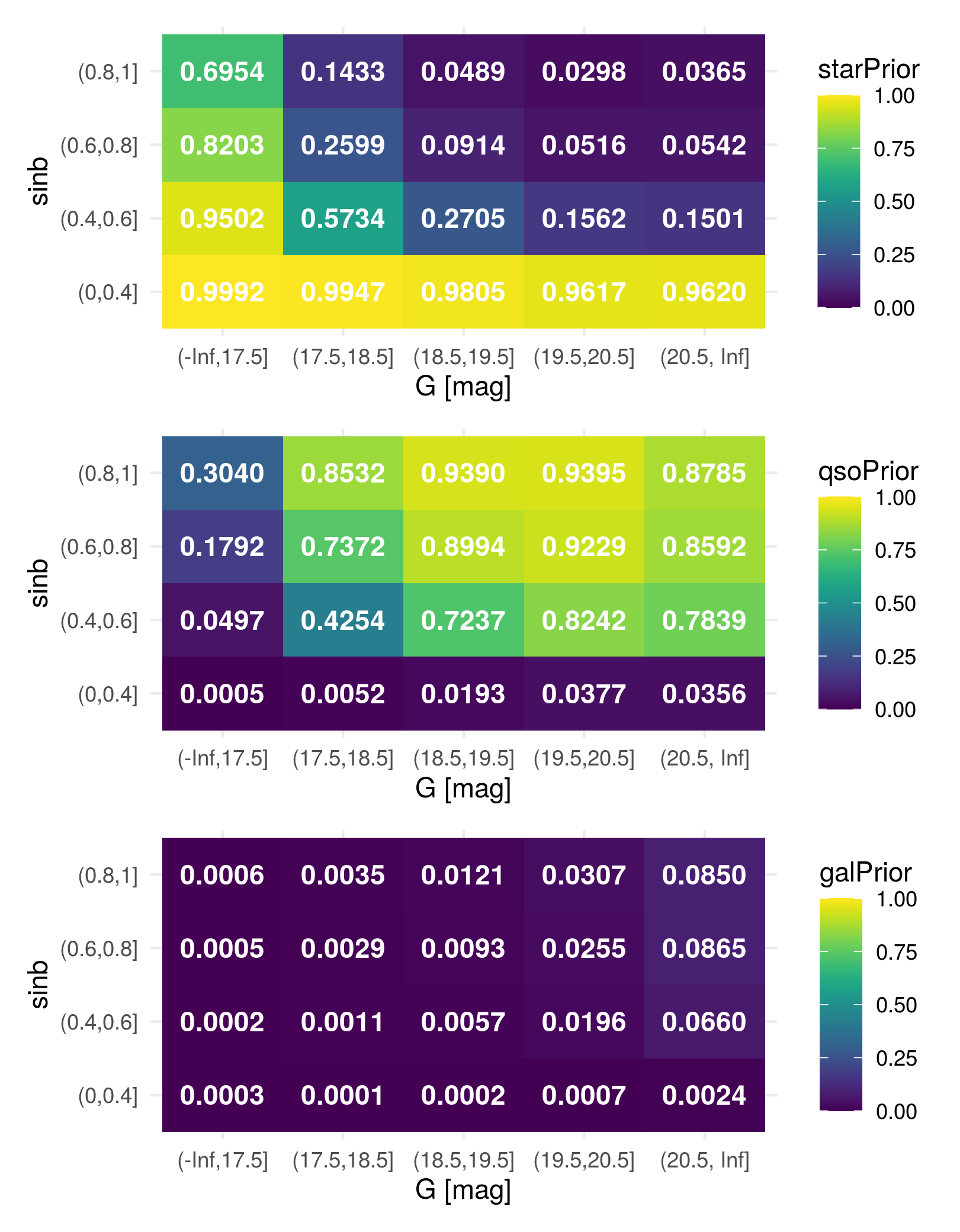}
\caption{Heatmap of the mixed prior distribution for the GDR3 Quasar Candidate Table. In this representation, the number of stars at lowest latitude exceeds the number of observed quasars and galaxies.  }
\label{Application GDR3 QSO- fig:Data-mixedpriors}
\end{figure}

\begin{figure}[]
\centering
\includegraphics[scale=0.7]{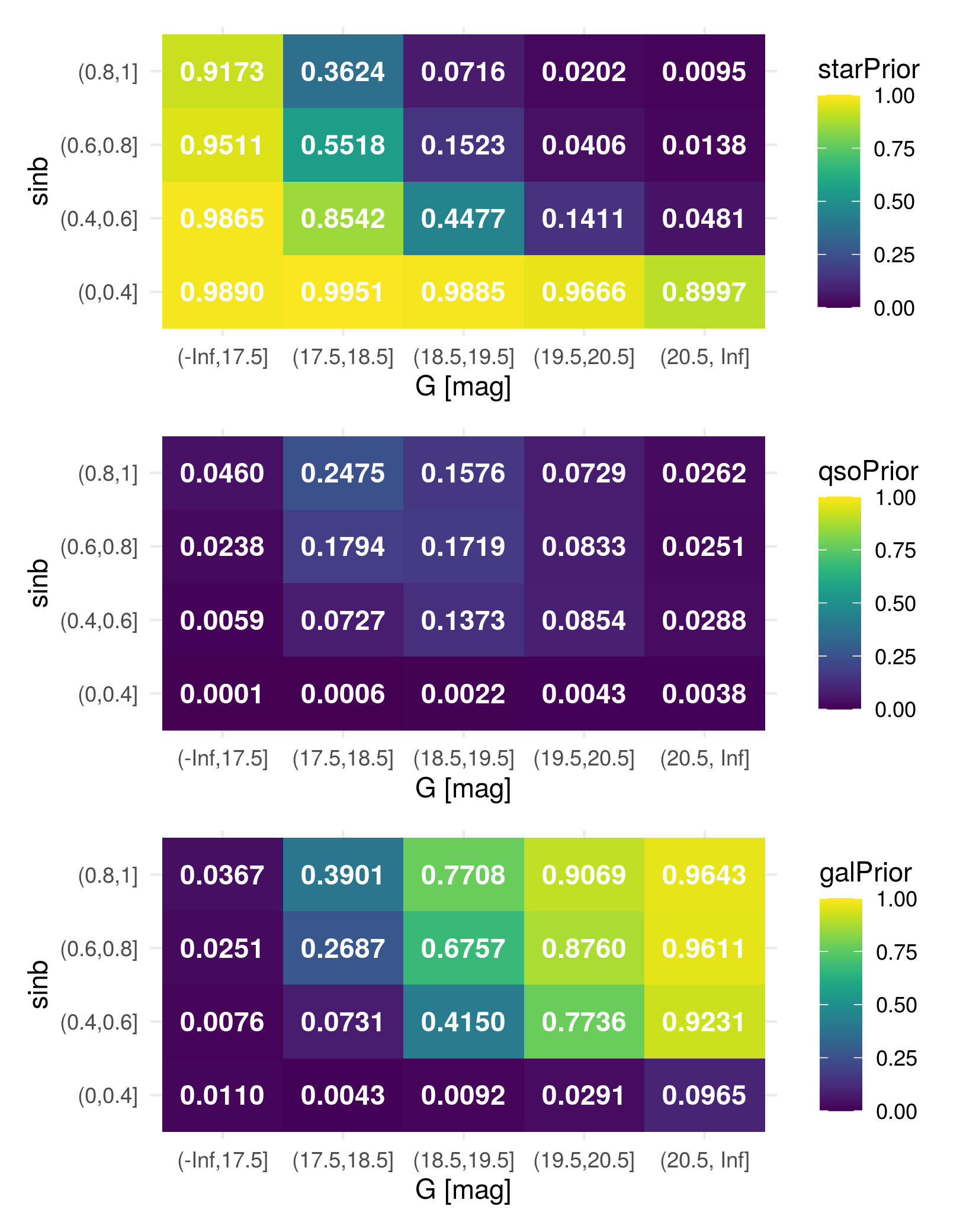}
\caption{Heatmap of the mixed prior distribution for the GDR3 Galaxy Candidate Table. In this representation, the number of stars at lowest latitude exceeds the number of observed quasars and galaxies.  }
\label{Application GDR3 Gal- fig:Data-mixedpriors}
\end{figure}

Our priors - both global and mixed - are designed for a sample of sources drawn at random from the Gaia/CatWISE2020 all sky sample. These priors are not appropriate for the classification of the GDR3 extragalactic tables in Sect.~\ref{sec:Quasar Candidates} and Sect.~\ref{Galaxy Candidates}, where we have 50-95\% extragalactic objects, rather than 0.1-0.2\% as expected by the prior. For application to these, we redefine our global priors by taking the purity of each GDR3 extragalactic table as defined in ~\citep{gaia_collaboration_gaia_2022}, which we denote as $p$. Considering the case of the quasar table, the normalised global prior becomes $(1-p-e,p,e)$, where $e$ is an estimation of the contamination from the galaxy class. The prior would be defined as $(1-p-e,e,p)$ in the case of the galaxy class. The normalised global priors are $(0.454,0.520,0.026)$  and $(0.274,0.036,0.690)$ , with the re-adjusted mixed priors shown in Fig.~\ref{Application GDR3 QSO- fig:Data-mixedpriors} and Fig.~\ref{Application GDR3 Gal- fig:Data-mixedpriors}  for the GDR3 quasar and galaxy candidate tables respectively.


\subsection{Application on a random subset of the overlap of GDR3 and CatWISE2020}
For the 50 million sources at the intersection of GDR3 and CatWISE2020, the true class of the source is unknown and therefore reliable performance metrics cannot be computed. However, we can compare the number of sources classified with the different priors, and compare the counts to expectations. We find  12607 quasars and  41153 galaxies or 1/4000 and 1/1200, using the global prior. When compared to the global prior values of 1/1000 for quasars and 1/500 for galaxies, we find that our results give a factor of 4 fewer quasars and a factor of 2 fewer galaxies.
Using the mixed prior, we find  97294 quasars and 192231  galaxies or 1/500 and 1/300. The mixed prior finds nearly 8 times as many quasars ($97294/12607=7.7$) and roughly 5 times more galaxies ($192231/41153=4.7$) as the global prior. This may be attributed to the mixed prior being very non-uniform in magnitude and latitude, similar to the true distribution: and by construction the mixed prior is better matched to the data. 

\begin{figure*}[]
\centering
\makebox[\textwidth][c]{\includegraphics[scale=0.6]{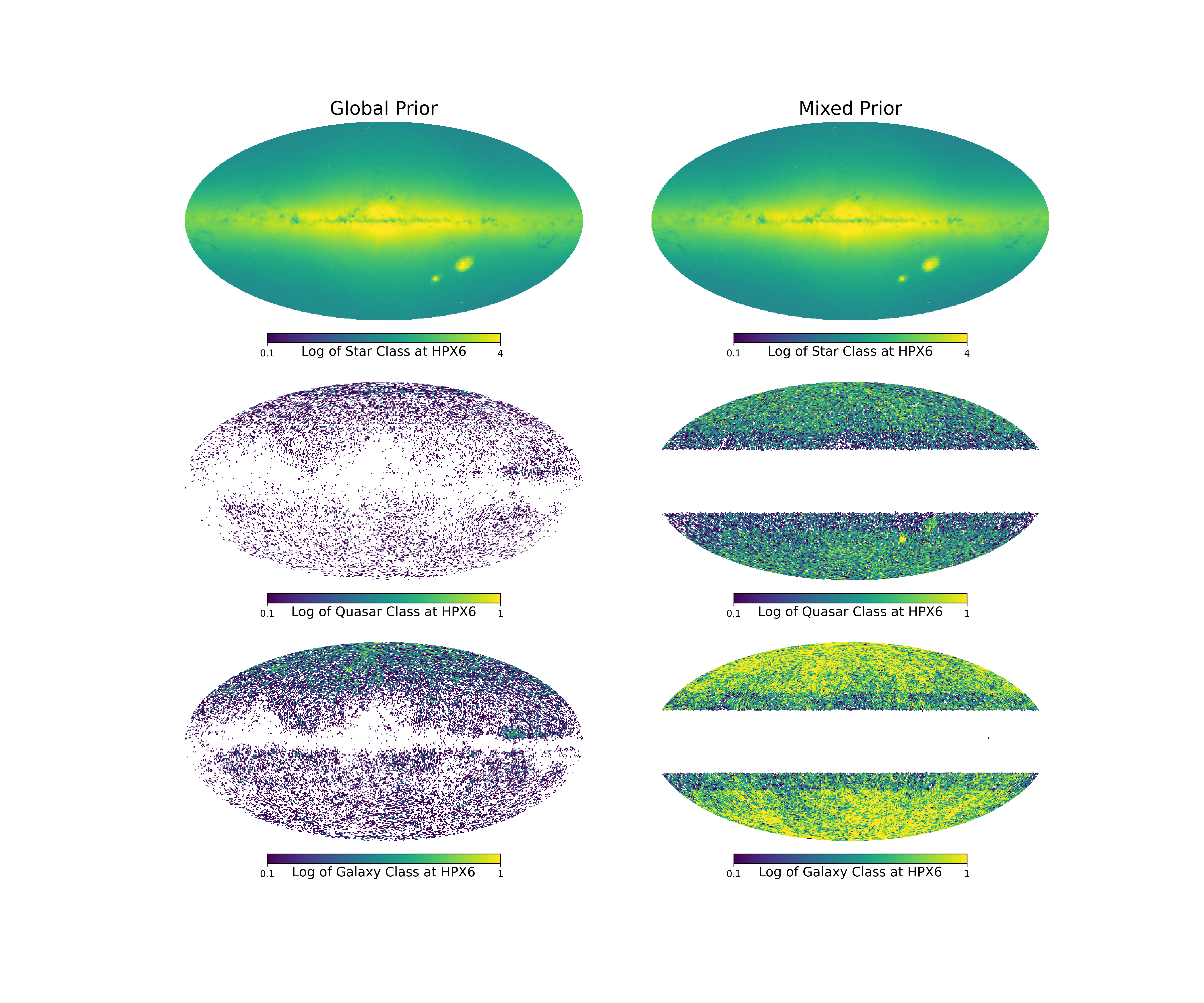}}
\caption{Log10 of counts for sources classified on the random Gaia DR3 sample on a healpix at level 6 (HPX6). As described in Sect.~\ref{mixedprior} the mixed prior is discretised by latitude and magnitude, this giving rise to the banded structure in the right panels. The white colour indicates a source density below the scale and anything above the scale is yellow.}
\label{fig:Discussion-3Class_combined_port_300dpi}
\end{figure*}

In Fig.~\ref{fig:Discussion-3Class_combined_port_300dpi}, we show the sky distributions of the sources by assigned class. For both priors, in a random sample of 50 million sources observed by Gaia, we classify less than 1\% of the sample as either a galaxy or a quasar, highlighting the scarcity of the extragalactic sources.

It is interesting to compare our results with those used from the DSC-Combmod classifier, which was used to identify many quasars and galaxies published in the GDR3 extragalactic candidates tables~\citep{gaia_collaboration_gaia_2022}. Combmod is the combination of the class posterior probabilities from two classifiers, DSC-Specmod and DSC-Allosmod~\citep{delchambre_gaia_2022}. Specmod classifies objects using BP/RP spectra, whereas Allosmod uses a GMM to classify objects using several astrometric and photometric features (the features being our Gaia\_f set, plus parallax and proper motion; see also~\citealt{bailer-jones_quasar_2019}). We use the quasar and galaxy class probabilities from Combmod, but take the star class probabilities to be one minus the sum of the quasar and galaxy probabilities (because Combmod reports more than three classes), and assign the class label to the class with the largest probability.
When applying the global prior, we identify 7\% of the Combmod quasars as quasars with the remaining 92.9\% identified as stars and 0.1\% as galaxies. We identify 21\% of the Combmod galaxies as galaxies, with the remaining 78.9\% identified as stars and 0.1\% as galaxies.
Using the mixed prior, we classify 40\% of the Combmod quasars as quasars with the remaining 59.8\% identified as stars and 0.2\% as galaxies. For the Combmod galaxies using the mixed prior, we find 56\% to be galaxies with the remaining 43.6\% as stars and 0.4\% as quasars.

\begin{figure*}[]
\centering
\includegraphics[scale=0.7]{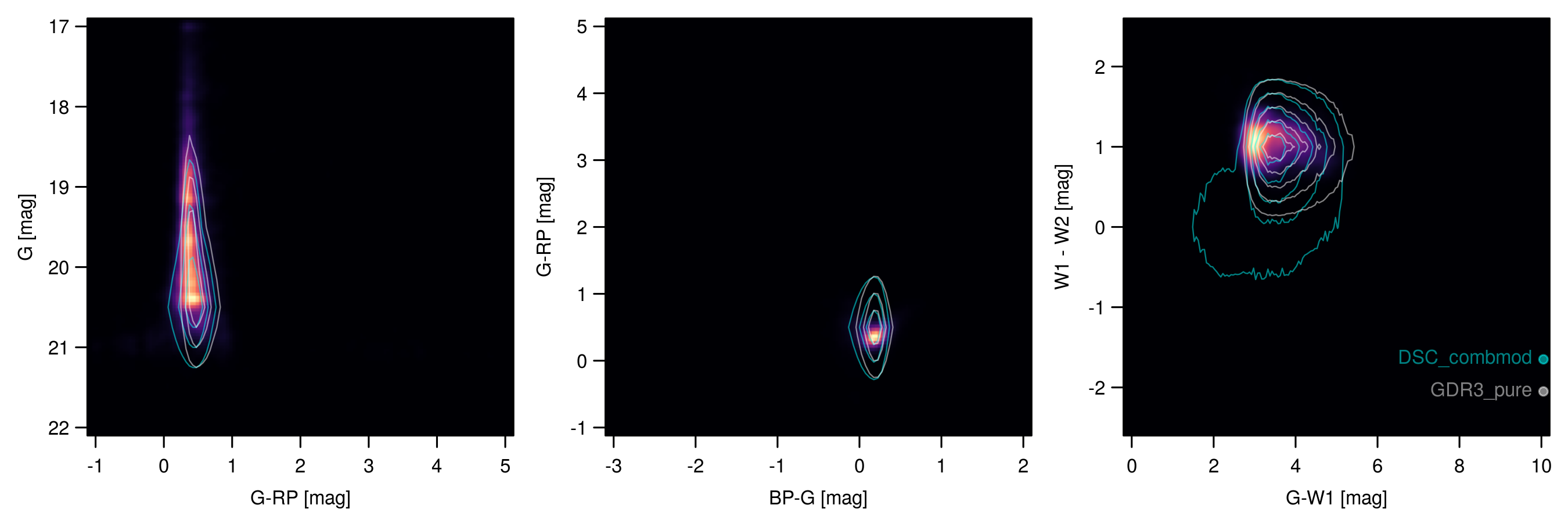}
\caption{Results on the randomly-selected set of 50 million Gaia DR3 and catWISE2020 sources using the global prior: Colour-magnitude and colour-colour diagrams for the quasars. Sources from the classifier adjusted by the global prior is given by the density scale where black is zero density and yellow is high density. DSC-Combmod sources are identified by the cyan contours and the GDR3 defined pure quasar sample by the white contours.
This colouring  will be used for the subsequent colour-magnitude and colour-colour diagrams.}
\label{fig:Random Gaia DR3_qso_combmod_colcol_GP}
\end{figure*}

\begin{figure*}[]
\centering
\includegraphics[scale=0.7]{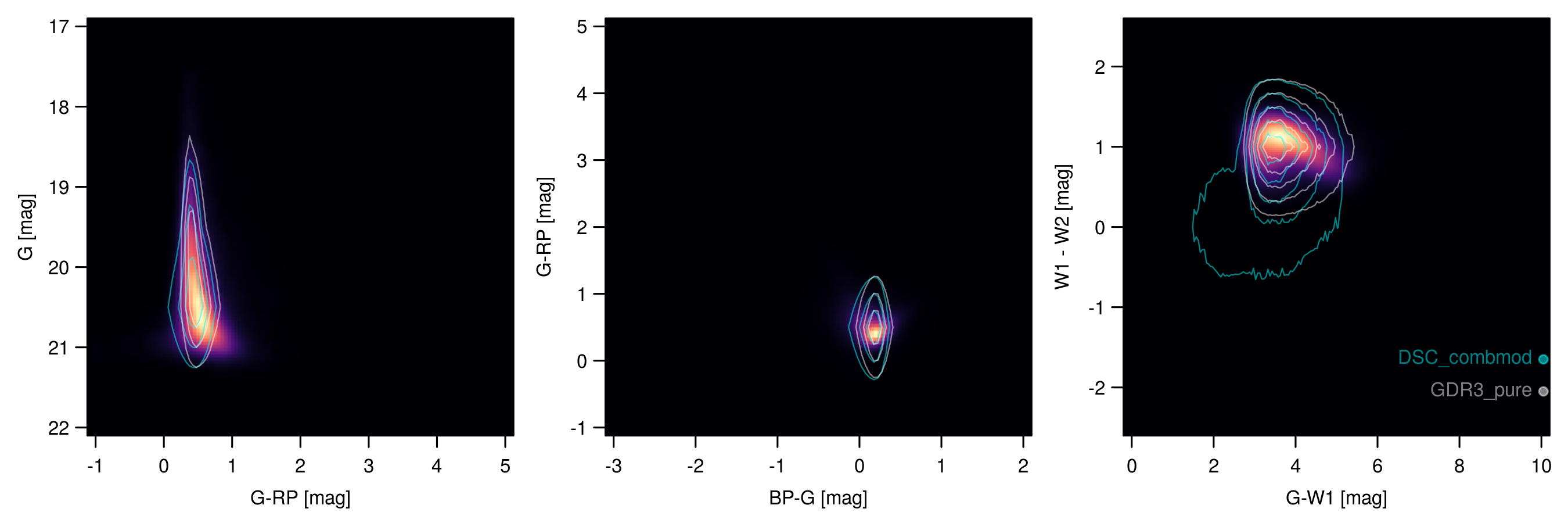}
\caption{Results on the randomly-selected set of 50 million Gaia DR3 and catWISE2020: As Fig.~\ref{fig:Random Gaia DR3_qso_combmod_colcol_GP} but using the mixed prior in our classifier.}
\label{fig:Random Gaia DR3_qso_combmod_colcol_MP}
\end{figure*}

\begin{figure*}[]
\centering
\includegraphics[scale=0.7]{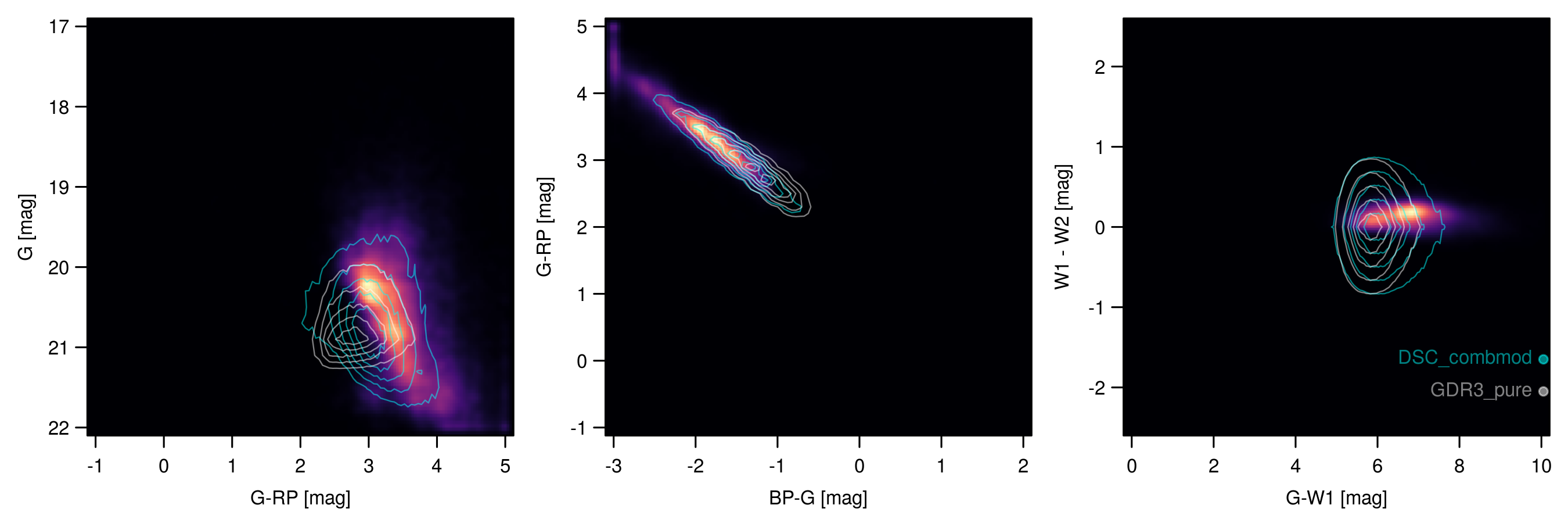}
\caption{Results on the randomly-selected set of 50 million Gaia DR3 and catWISE2020: Colour-magnitude \& Colour-colour diagrams for the galaxies derived from DSC-Combmod, the GDR3 defined pure galaxy sample and from the classifier adjusted by the global prior.}
\label{fig:Random Gaia DR3_gal_combmod_colcol_GP}
\end{figure*}

\begin{figure*}[]
\centering
\includegraphics[scale=0.7]{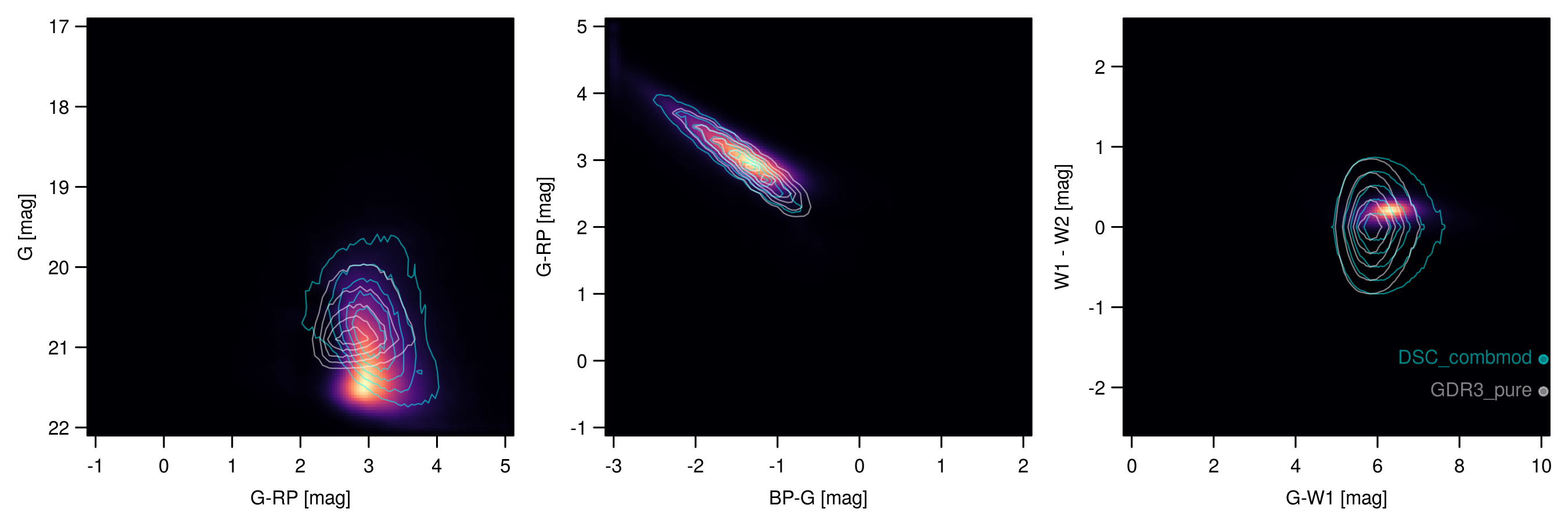}
\caption{Results on the randomly-selected set of 50 million Gaia DR3 and catWISE2020: Colour-magnitude \& Colour-colour diagrams for the galaxies using DSC-Combmod, the GDR3 defined pure galaxy sample and the classifier adjusted by the mixed prior.}
\label{fig:Random Gaia DR3_gal_combmod_colcol_MP}
\end{figure*}

We now refine the 50 million sources by considering those that are classified as a quasar or a galaxy in the pure samples defined in the GDR3 quasar and galaxy candidate tables respectively. We aim to see whether the proportion of identified quasars and galaxies increases, when the sample is refined. We find that our classifier identifies 12\% of the quasars in the pure quasar candidate table using the the global prior, an improvement of 5\% compared to quasars classified in Combmod. Using the mixed prior, we identify 69\% of the quasars in the pure quasar candidate table, over 25\% better than the Combmod quasars. When considering the pure galaxy candidate table, we identify 18\% as galaxies using the global prior which is a reduction of 3\% when compared. A 2\% reduction is seen when applying the mixed prior, where we identify 54\% of galaxies in the pure galaxy candidate table.
Using the three different classifications -- DSC-Combmod, the pure subsample from the GDR3 candidate table, and our classifier -- we illustrate the density of the predicted sources in colour-colour diagrams and a colour-magnitude diagram, with the contours representing the classifications from DSC-Combmod and the purer subsamples.
Figure~\ref{fig:Random Gaia DR3_qso_combmod_colcol_GP} shows the sources classified as quasars, using the global prior in our classifier. We see that considerably fewer sources are classified as extragalactic when using the global prior compared to the mixed prior in Fig.~\ref{fig:Random Gaia DR3_qso_combmod_colcol_MP}, but are focused towards the redder magnitude. In contrast to the global prior, the mixed prior allows for more freedom in the identification of sources that are quasars, closely resembling the contours of the pure sample.
Figure~\ref{fig:Random Gaia DR3_gal_combmod_colcol_GP} and Fig.~\ref{fig:Random Gaia DR3_gal_combmod_colcol_MP} represent the density of the galaxy class with the global prior and mixed prior adjustment respectively. The global prior results are a subset of the mixed prior, with the mixed prior extending to bluer G-RP but the global prior not extending redder.


\begin{figure}[]
\centering
\includegraphics[scale=0.7]{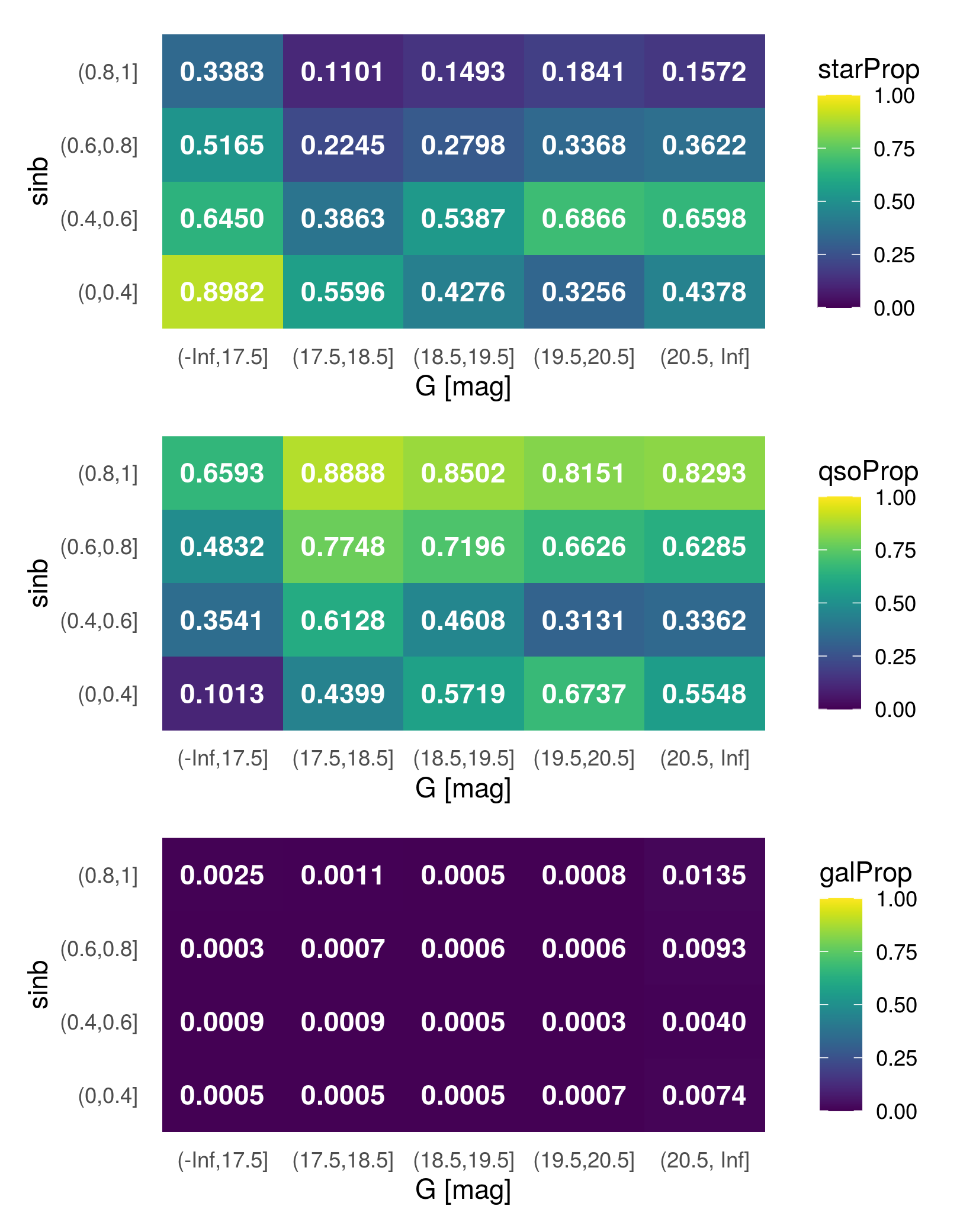}
\caption{Heatmap of the distribution for quasars identified using the global prior for the GDR3 Quasar Candidates as function of magnitude and latitude. Each mag/lat cell is normalised across the three classes.}
\label{Application GDR3 QSO- fig:Data-heatmapGP}
\end{figure}


\begin{figure}[]
\centering
\includegraphics[scale=0.7]{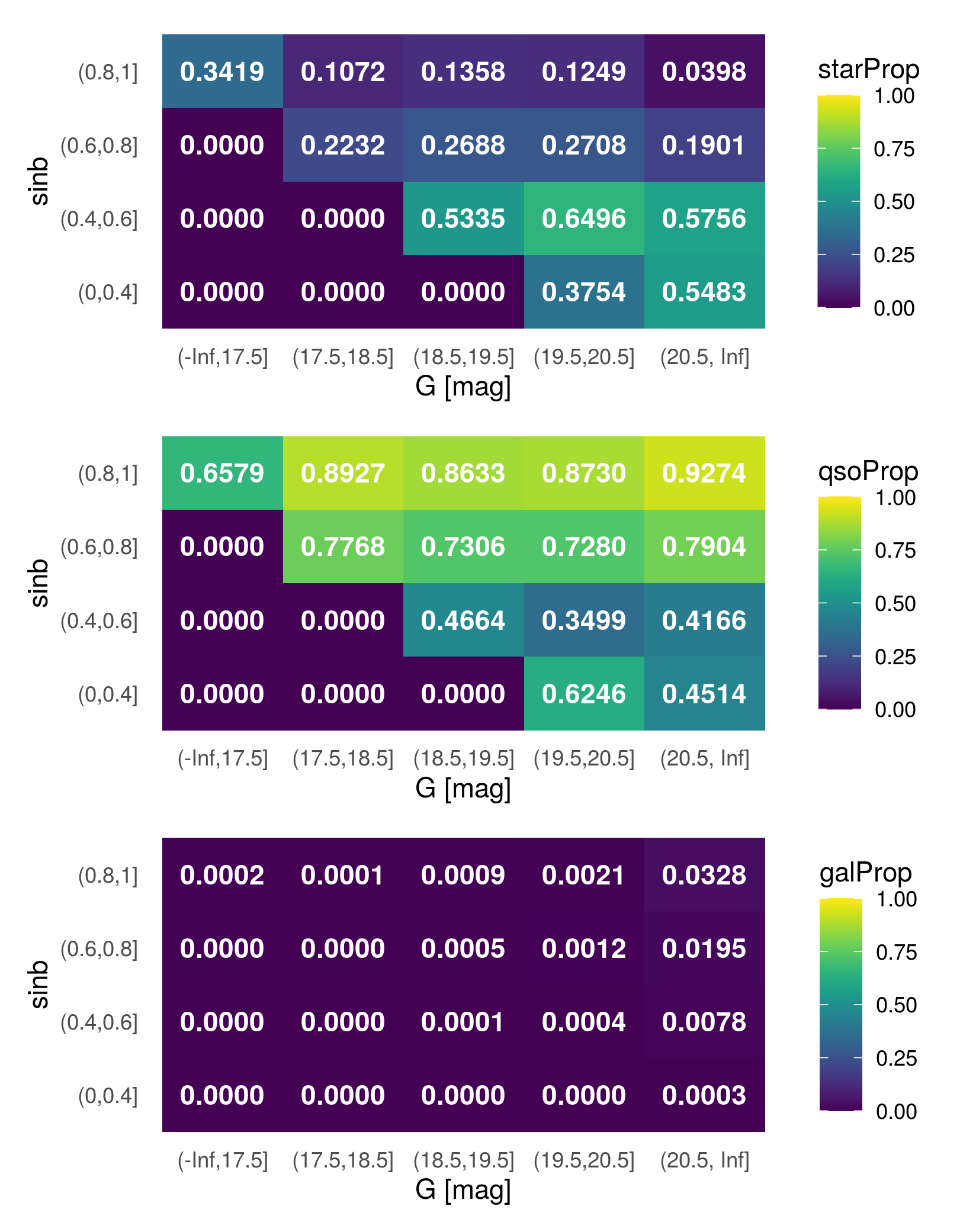}
\caption{Heatmap of the distribution for quasars as in Fig.~\ref{Application GDR3 QSO- fig:Data-heatmapGP} but using the mixed prior in our classifier.} 
\label{Application GDR3 QSO- fig:Data-heatmapMP}
\end{figure}

\begin{figure*}[]
\centering
\includegraphics[scale=0.7]{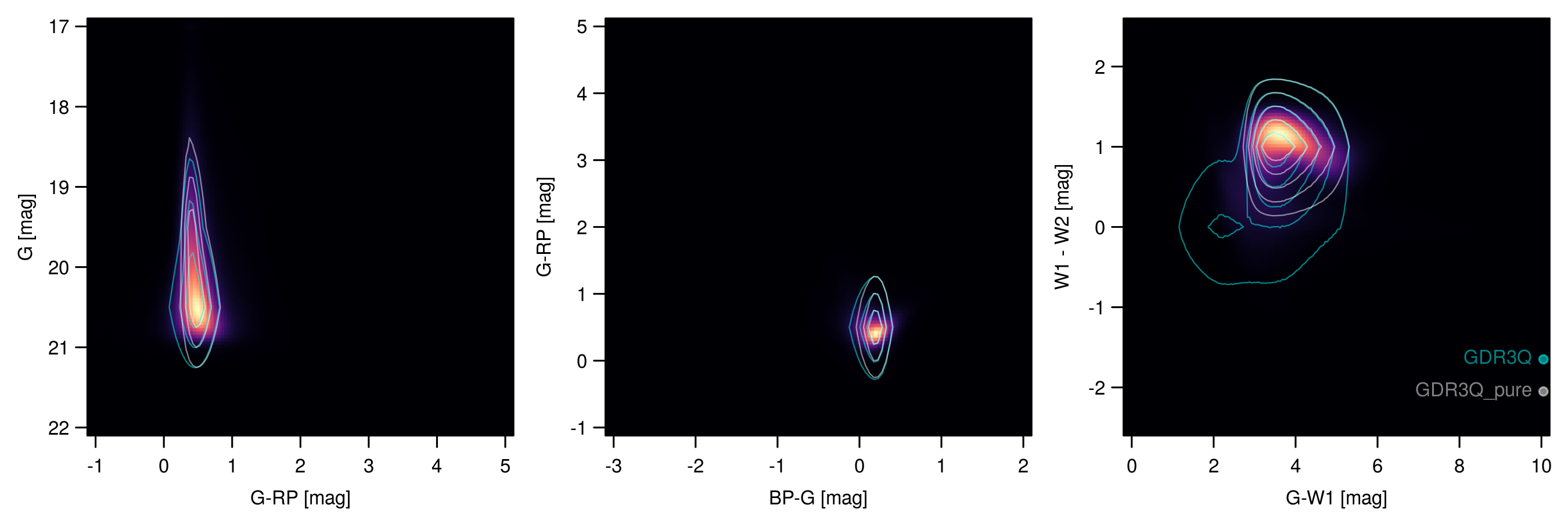}
\caption{GDR3 Quasar Candidate Table Mixed Prior: Colour-magnitude \& Colour-colour diagrams for the quasars using the mixed prior. The sources identified by the classifier are represented by the density scale, where black is zero density and yellow is high density. 
GDR3 quasar sources are identified by the cyan contours and the GDR3 pure quasar sample by the white contours.}
\label{fig:GDR3 Quasar_qso_combmod_colcol_MP}
\end{figure*}

\begin{figure*}
     \begin{minipage}[t]{.55\linewidth}
		\includegraphics[width=0.85\textwidth, page=1]{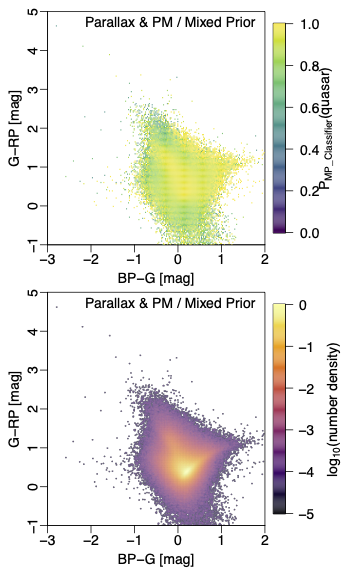}  
      \end{minipage}
      \begin{minipage}[t]{0.55\linewidth}
		\includegraphics[width=0.85\textwidth, page=1]{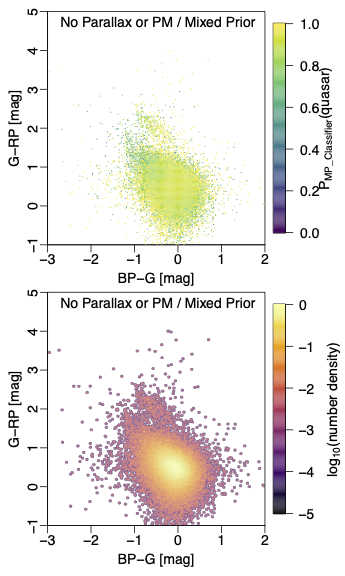}  
      \end{minipage}
      \captionof{figure}{GDR3 Quasar Candidate Table Mixed Prior: Probability and density distributions for sources classified as a quasar. The left hand side panels correspond to sources with parallax while the right hand side panels represent the distribution for sources without parallax.}  
      \label{fig:GDR3 Quasar Mixed Prior: Probability and density distributions }
\end{figure*}


\begin{figure}[]
\centering
\includegraphics[scale=0.7]{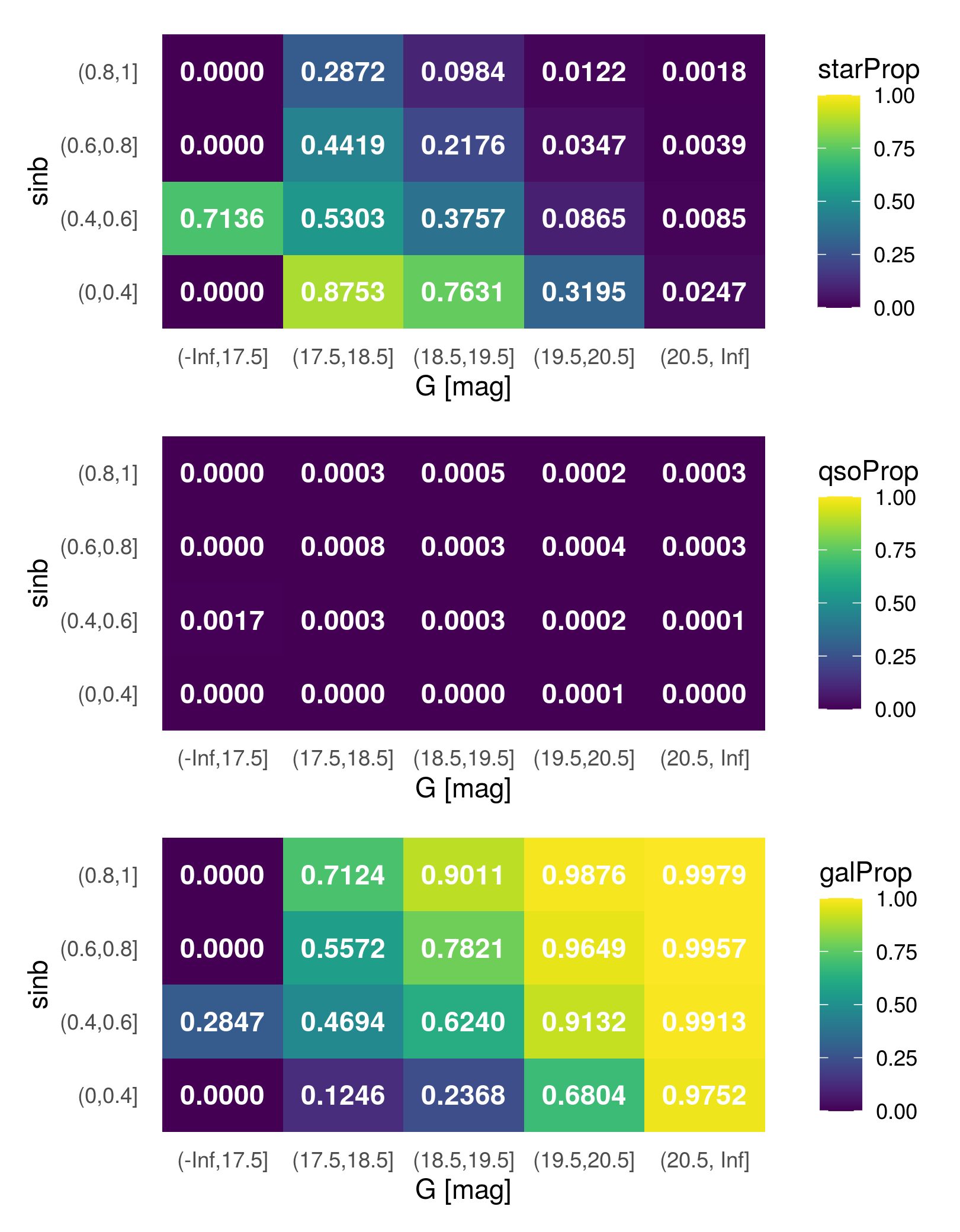}
\caption{Heat map of the distribution for galaxies identified using the global prior for the GDR3 Galaxy Candidates by magnitude and latitude.}
\label{Application GDR3 Gal- fig:Data-heatmapGP}
\end{figure}


\begin{figure}[]
\centering
\includegraphics[scale=0.7]{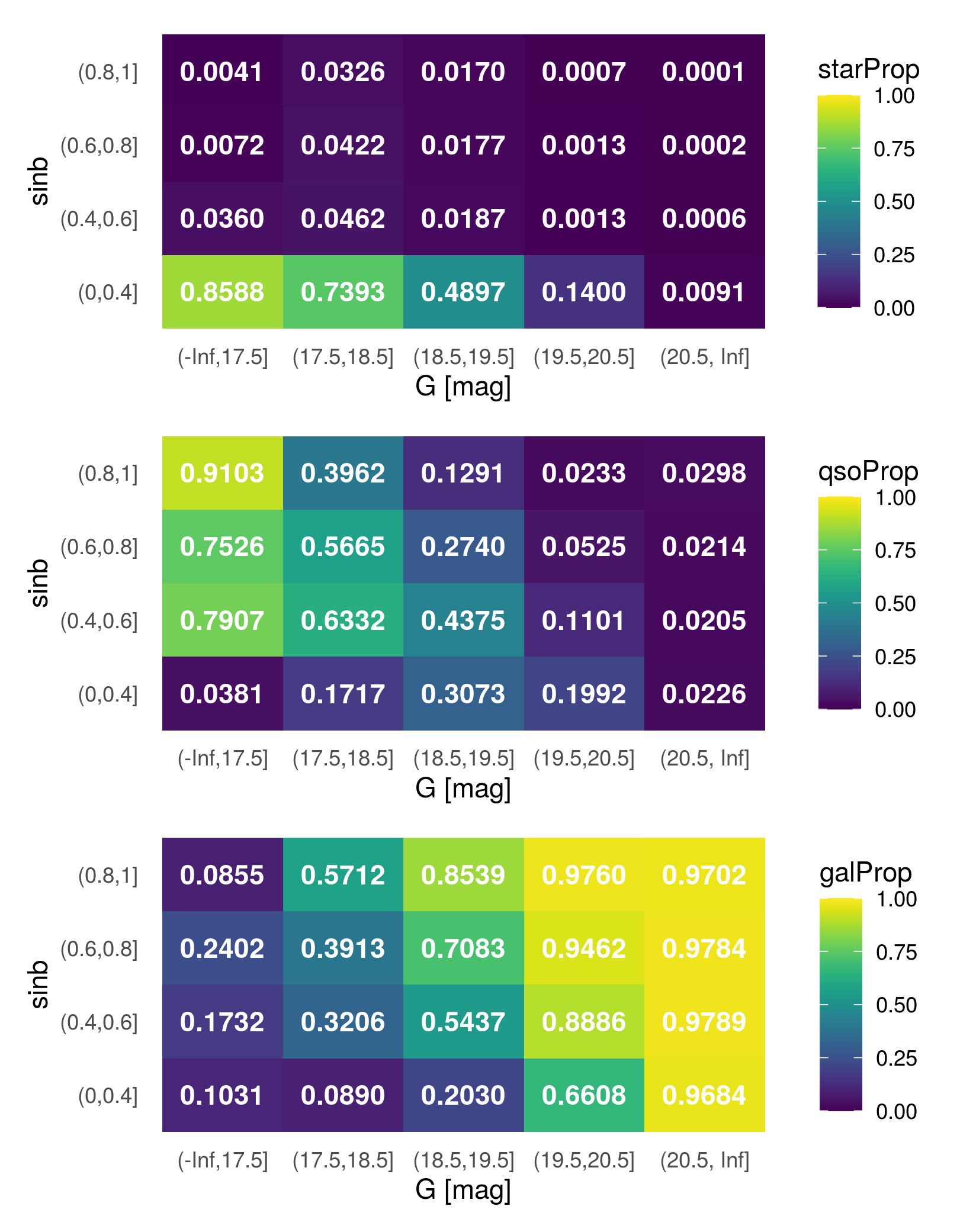}
\caption{Heat map of the distribution for galaxies as in Fig.~\ref{Application GDR3 Gal- fig:Data-heatmapGP} but using the mixed prior in our classifier.}
\label{Application GDR3 Gal- fig:Data-heatmapMP}
\end{figure}

\begin{figure*}[]
\centering
\includegraphics[scale=0.7]{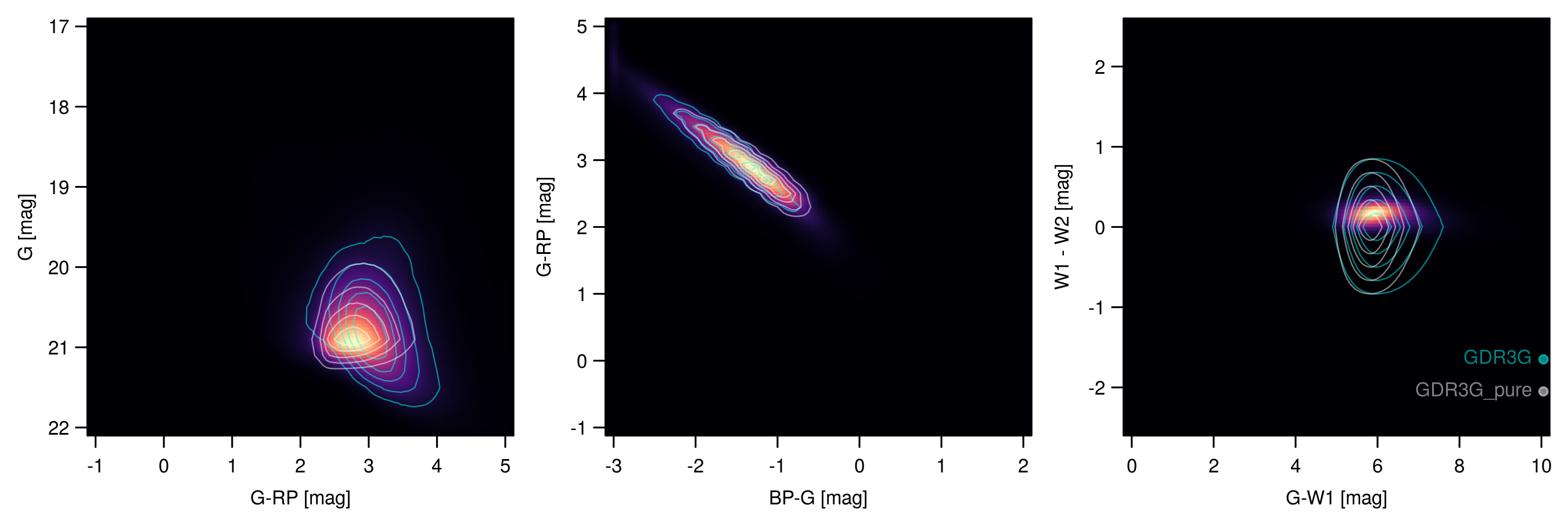}
\caption{GDR3 Galaxy Candidate Table Mixed Prior: Colour-magnitude \& Colour-colour diagrams for the galaxies using the mixed prior. The sources identified by the classifier are represented by the density scale, where black is zero density and yellow is high density. 
GDR3 galaxy sources are identified by the cyan contours and the GDR3 pure galaxy sample by the white contours.}
\label{fig:GDR3 Galaxy_gal_combmod_colcol_MP}
\end{figure*}


\begin{figure*}
     \begin{minipage}[t]{.55\linewidth}
		\includegraphics[width=0.85\textwidth, page=1]{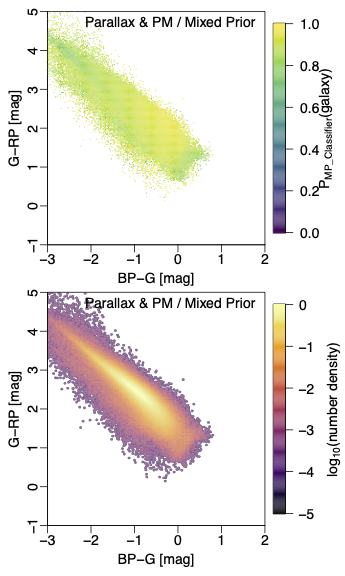}  
      \end{minipage}
      \begin{minipage}[t]{0.55\linewidth}
		\includegraphics[width=0.85\textwidth, page=1]{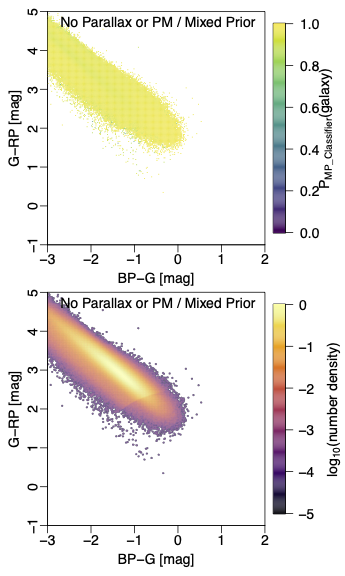}  
      \end{minipage}
      \captionof{figure}{GDR3 Galaxy Candidate Table Mixed Prior: Probability and density distributions for sources classified as a galaxy The right hand side panels correspond to sources with parallax while the left hand side panels represent the distribution for sources without parallax.We find in the bottom-right panel a similar colour excess factor locus at BP-G =-0.5 and G-RP=2, as in figure 3 of \cite{bailer-jones_quasar_2019} and figure 31 of \cite{gaia_collaboration_gaia_2022}. This locus is however not evident in the case which has parallax and proper motions.}   
      \label{GDR3 Galaxy Mixed Prior: Probability and density distributions}
\end{figure*}

\begin{table*}
\small
\centering
\caption{Quasar Candidates: Counts of objects in the predicted classes and the proportion identified as quasars using the extragalactic table tuned prior defined in Sect.~\ref{Sec:Application}. GP and MP refer to the global prior and mixed prior respectively.}
\begin{tabular}{clllll}
    & & \multicolumn{1}{c}{Predicted} \\
    \toprule
     & & Star & Quasar & Galaxy & Quasar proportion \\ \midrule
    \multirow{4}{*}{\rotatebox{90}{GP}}
        & GDR3 Quasar                   & 1826019   & 2211696    & 10911   &  0.5463   \\  
        & Pure GDR3 Quasar              & 54006     & 1768694    &  222   & 0.9703    \\  
        & SDSS16 Quasar + GDR3          & 753       & 401104     &  5    & 0.9981    \\ 
        & SDSS16 Quasar + Pure GDR3     & 725       & 394418     &  5    & 0.9982    \\ 
    \bottomrule 
    \multirow{4}{*}{\rotatebox{90}{MP}}
        & GDR3 Quasar                   &  1656379  & 2372430   & 19817  & 0.5860    \\ 
        & Pure GDR3 Quasar              & 68680     & 1753917   &  325  & 0.9621   \\ 
        & SDSS16 Quasar + GDR3          & 491       & 401369    &  2   & 0.9988    \\ 
        & SDSS16 Quasar + Pure GDR3     & 466       & 394680    &  2   & 0.9988    \\ 
    \bottomrule 

\end{tabular}
\label{cm:QSO Set - updated prior}
\end{table*}

\subsection{Application to quasar candidates from GDR3}\label{sec:Quasar Candidates}
The GDR3 quasar candidate table defined in \cite{gaia_collaboration_gaia_2022} contains 6.6 million potential quasars with a purity of 52\%, and is further refined into a pure sub-sample containing 1.9 million quasars with a purity of 95\%. The overlap with CatWISE2020 results in 4\,048\,626 GDR3 quasars and 1\,822\,922 pure sub-sample quasars.

We applied our trained classifier from Sect.~\ref{Subsec:Max Source Classifier} to the GDR3 quasar candidates overlap with CatWISE2020 and estimate the probabilities of the three classes. We assess the classification performance of our model by considering the proportion of quasars identified by our classifier using the global prior and the mixed prior re-defined for this application dataset (as explained at the beginning of this section), on the assumption that the quasar candidate overlap is entirely quasars.

The results are shown in Table~\ref{cm:QSO Set - updated prior}, where in the global prior case we identify 55\% of quasars in the GDR3 candidate table. If we further constrain the sample by considering the pure sub-sample only or in the pure sub-sample and in the SDSS16 quasar table, we see the proportion of quasars identified by our classifier is considerably higher than the GDR3 candidate sample, at 99.8\%. A similar trend is reported in the mixed prior case, however identifying 58\% of quasars in the GDR3 candidate table and 99.9\% when restricting the sample to be the pure sub-sample or the pure sub-sample and the SDSS16 quasar table.
Given that both global prior and mixed prior have the same global prior behind them, adding the highly non-uniform distribution of the latitude/G dependence to the prior makes it slightly more favourable to finding quasars and galaxies where we expect to find them.
We can deconstruct this results table further by considering the entire GDR3 quasar candidate sample in Fig.~\ref{Application GDR3 QSO- fig:Data-heatmapGP} for the global prior and Fig.~\ref{Application GDR3 QSO- fig:Data-heatmapMP} for the mixed prior. Comparing the two priors we see a higher proportion of quasars identified in the fainter and higher magnitude end in the mixed prior case than the global prior, but the distribution on average is quite similar.

We visualise the application of the mixed prior to the quasar candidate table in Fig.~\ref{fig:GDR3 Quasar_qso_combmod_colcol_MP}, and the considerable overlap between the pure sample contours in the most dense region of the mixed prior classifier is evident. The same distribution can be seen in the case of the global prior. 

By splitting the sample into two subsets based on the availability of parallax or proper motions in Fig.~\ref{fig:GDR3 Quasar Mixed Prior: Probability and density distributions }, we observe a higher density distribution for sources with parallaxes compared to the sources without parallax measurements. Furthermore, for the sources classified with parallax we see a shift in the density of the colour distribution, with more sources extending to $BP-G = 1$, whereas the sources without parallax and proper motions are centred around $BP-G = 0$ with a few outliers when $G-RP > 2$ for sources without parallax or proper motions in the case of the global prior. For the mixed prior the distribution in colour space is similar, however in the top-left panel for sources with $G-RP > 2$ the probabilities are less than in the case of the global prior. 
Overall, the application of our classifier to the quasar candidates from GDR3 identifies 96-97\% of the pure quasar sub-sample as quasars. Moreover, when requiring the source to have an an SDSS16 quasar classification, we identify 99.9\% of them as quasars, irrespective of whether the source was in the GDR3 pure sub-sample or not.

\subsection{Application to galaxy candidates from GDR3}\label{Galaxy Candidates}
Analogous to Sect.~\ref{sec:Quasar Candidates}, we here apply our classifier to the galaxy candidate table in GDR3, which comprises 4.8 million candidates with a purity of 69\%, and includes a purer sub-sample of 2.8 million candidates with a purity of 94\%.  The overlap with CatWISE2020 reduces the counts to 4\,194\,100  and 2\,824\,570 respectively.

From Table~\ref{cm:Gal Set - Updated prior} we find the proportion of galaxies identified by our classifier in the full galaxy candidate table to be 93\% when using the global prior and if we apply the mixed prior to this table we find 91\%. If we further constrain the sample by considering the pure sub-sample or in the pure sub-sample and in the SDSS16 galaxy table, we see the proportion of galaxies identified by our classifier is higher, at 99\% for the both priors.
Exploring the entire GDR3 galaxy candidate sample further in Fig.~\ref{Application GDR3 Gal- fig:Data-heatmapGP} for the global prior and Fig.~\ref{Application GDR3 Gal- fig:Data-heatmapMP} for the mixed prior. Comparing the two priors we see a higher proportion of galaxies identified in the fainter and higher magnitude end in the global prior case than the mixed prior, but the distribution on average is quite similar. Furthermore the mixed prior considers more galaxy sources to be quasars particularly at the bright end and higher latitudes, whereas the global prior considers more galaxy sources to be stars at the bright end but lower latitudes.

We can see this distribution for the mixed prior results in Table~\ref{cm:Gal Set - Updated prior} and in Fig.~\ref{fig:GDR3 Galaxy_gal_combmod_colcol_MP}. We see closer contours for the GDR3 pure sample centred around the highest density region when using the mixed prior classifier and wider contours for the GDR3 sample as expected. A similar result is seen when applying the global prior.
In contrast to the work by ~\cite{bailer-jones_quasar_2019}, our classifier was fit without using parallax or proper motions, in order to retain as many galaxy sources as possible. We assess in Fig.~\ref{GDR3 Galaxy Mixed Prior: Probability and density distributions} whether our classifier has a different distribution in either the count or probability spaces for sources with parallax and proper motions and for those that do not. We see for the sources classified using the mixed prior without parallax and proper motions a tendency towards redder magnitudes. The probability distributions are unperturbed and follow a similar trend with higher probabilities towards the lower magnitudes.

\begin{table*}
\small
\centering
\caption{Galaxy Candidates: Counts by predicted class and proportion identified as galaxies using the extragalactic-table-tuned prior defined in Sect.~\ref{Sec:Application}. GP and MP refer to the global prior and mixed prior respectively.}
\begin{tabular}{clllll}
\multicolumn{5}{c}{Predicted} \\
    \toprule
    & & Star & Quasar & Galaxy & Galaxy proportion \\ \midrule
    \multirow{4}{*}{\rotatebox{90}{GP}}
        & GDR3 Galaxy                   & 306073   & 913     & 3887114    & 0.9268   \\ 
        & GDR3 Pure Galaxy              & 1834   & 20     & 2822716    & 0.9993   \\ 
        & SDSS16 Galaxy + GDR3          & 27    & 3     &  514735   & 0.9999  \\ 
        & SDSS16 Galaxy + Pure GDR3     & 1    & 0     &  393043   & 1.0000   \\ 
    \bottomrule 
    \multirow{4}{*}{\rotatebox{90}{MP}}
        & GDR3 Galaxy                   &  128638  &     264534  & 3800928   & 0.9062   \\ 
        & GDR3 Pure Galaxy              & 710   &    41149    & 2782711   & 0.9852   \\ 
        & SDSS16 Galaxy + GDR3          & 3     &    323     & 514439    & 0.9994   \\ 
        & SDSS16 Galaxy + Pure GDR3     & 0     &    151      & 392893    & 0.9996   \\ 
    \bottomrule 
\end{tabular}
\label{cm:Gal Set - Updated prior}
\end{table*}

\textbf{
\begin{table*}
\small
\centering
\caption{Quasar Candidates: A subset of the table of mixed prior probabilities as calculated on the Quasar candidate table from GDR3. The full tables of probabilities as calculated on the Quasar candidate table from GDR3 and on the Galaxy candidate table from GDR3 are available upon request.}
\begin{tabular}{llllll}
 \toprule
source\_id & isQSO\_pure & isQSO\_SDSS & pStar & pQSO & pGAL \\ \midrule
 3470333738112& 1 & 1 & 0.0001936 & 0.9998064 & 0.0000000 \\
 5944234902272& 1 & 1 & 0.0001846 & 0.9998154 & 0.0000000 \\
 6459630980096& 1 & 0 & 0.0009402 & 0.9969757 & 0.0020841 \\
 9517648372480& 1 & 0 & 0.0001880 & 0.9998120 & 0.0000000\\
 10655814178816& 1 & 0 &0.0001485 & 0.9998457 & 0.0000058\\
 $\cdots$ & $\cdots$ & $\cdots$ & $\cdots$ &$\cdots$ & $\cdots$\\
 \bottomrule 
\end{tabular}
\label{probs:sample}
\end{table*}}


\section{Conclusions}
Building large catalogues of well-classified extragalactic sources is useful for large-scale statistical analyses in astronomy. 
In this paper we look at how adding infrared improves the classification of extragalactic sources compared to just using Gaia. Our results indicate an improved classification performance when adding the infrared colour information from CatWISE2020. The purities of the quasar and galaxy class improve from 0.9091 and 0.9759 to 0.9705 and 0.9784 respectively.
We discuss how using a prior and adjusting the confusion matrix to reflect the expected (high) level of stellar contamination in a real application are necessary steps in ensuring that the reported results are representative of what a classifier's performance will be when test or application datasets do not reflect the true class distribution.
Significantly, we find that using a prior that varies with latitude and magnitude gives higher purity {\em and} completeness for extragalactic objects: Looking at Fig.~\ref{fig:Results-3class_quasar_purity_mixedprior} in the adjusted case, and taking the bin where sinb = $(0.6,0.8]$ and G = $(18.5,19.5]$, we observe an improvement in the purity of the quasar class from 0.51 to 0.58. This result is coupled with a higher completeness seen in Fig.~\ref{fig:Results-3class_completeness_mixedprior}, from 0.84 to 0.97 in this bin. The published probabilities for the mixed prior classifier applied to the quasar and galaxy extragalactic candidate tables are available upon request.  Table~\ref{probs:sample} illustrates the format of the tables. 
Exploiting the results of our classifications would be useful to scientific studies focusing on extragalactic sources as well as investigating stellar populations in the Milky Way as observed by Gaia and CatWise2020.
Finally, when testing different statistical models we find that decision tree based methods, in particular XGboost, are more effective than Gaussian mixture models for this type of classification task.

\newpage
\begin{acknowledgements}
We thank Morgan Fouesneau and Rene Andrae for their useful comments that helped to improve this work. 
We also thank the referee for their suggestions and helpful comments.
The modelling and plots were done using \texttt{R} (\url{http://www.r-project.org}) and \texttt{Python} (\url{https://www.python.org}).
This research made use of the cross-match service provided by CDS, Strasbourg.
This work was funded in part by the DLR (German space agency) via grant 50 QG 2102.
Authors report the use of data from the European Space Agency (ESA) mission Gaia (\url{http://www.cosmos.esa.int/gaia}), processed by the Gaia Data Processing and Analysis Consortium (DPAC, \url{http://www.cosmos.esa.int/web/gaia/dpac/consortium}). Funding for the DPAC has been provided by national institutions, in particular the institutions participating in the Gaia Multilateral Agreement. 

\end{acknowledgements}

%
%

\bibliographystyle{aa} 
\bibliography{references.bib}


\begin{appendix} 

\section{Prior counts and latitude and magnitude prior}
The following section shows the the counts of sources in SDSS16 as a function of latitude and magnitude as well as the distribution of the prior. 

\begin{table} 

\centering
\begin{tabular}[t]{l|l|r|r|r}
\hline
 cutSinb & cutgMag & starN & qsoN & galN\\
\hline
(0,0.4] & (17.5,18.5] & 67338 & 184 & 81\\
\hline
(0,0.4] & (18.5,19.5] & 107892 & 1112 & 277\\
\hline
(0,0.4] & (19.5,20.5] & 159394 & 3274 & 1320\\
\hline
(0,0.4] & (20.5, Inf] & 130105 & 2526 & 3839\\
\hline
(0,0.4] & (-Inf,17.5] & 81827 & 22 & 250\\
\hline
(0.4,0.6] & (17.5,18.5] & 5644 & 2196 & 133\\
\hline
(0.4,0.6] & (18.5,19.5] & 8077 & 11332 & 2060\\
\hline
(0.4,0.6] & (19.5,20.5] & 11479 & 31767 & 17322\\
\hline
(0.4,0.6] & (20.5, Inf] & 10052 & 27529 & 53069\\
\hline
(0.4,0.6] & (-Inf,17.5] & 9921 & 272 & 21\\
\hline
(0.6,0.8] & (17.5,18.5] & 2910 & 4328 & 390\\
\hline
(0.6,0.8] & (18.5,19.5] & 4147 & 21412 & 5063\\
\hline
(0.6,0.8] & (19.5,20.5] & 6037 & 56641 & 35817\\
\hline
(0.6,0.8] & (20.5, Inf] & 5946 & 49410 & 113906\\
\hline
 (0.6,0.8] & (-Inf,17.5] & 5508 & 631 & 40\\
\hline
 (0.8,1] & (17.5,18.5] & 1928 & 6022 & 571\\
\hline
 (0.8,1] & (18.5,19.5] & 2625 & 26410 & 7771\\
\hline
(0.8,1] & (19.5,20.5] & 4227 & 69930 & 52335\\
\hline
(0.8,1] & (20.5, Inf] & 5214 & 65828 & 145758\\
\hline
(0.8,1] & (-Inf,17.5] & 3542 & 812 & 39\\
\hline
\end{tabular}
\caption{Prior Table Counts}
\label{Table:3Classs - Prior Counts}
\end{table}

\begin{figure}[]
\centering
\includegraphics[scale=0.7]{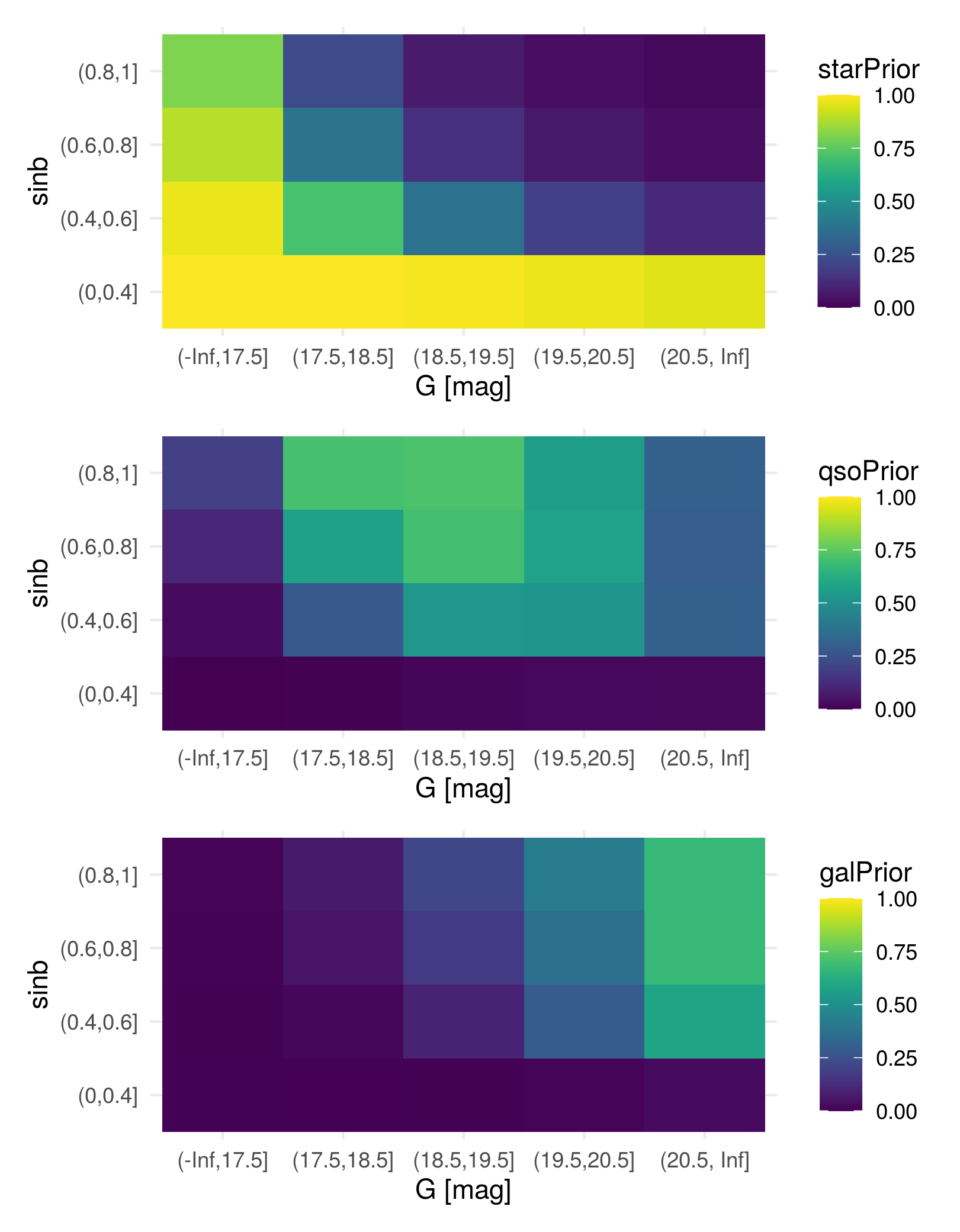}
\caption{Heatmap of the joint latitude and magnitude prior for each class. The top-panel refers to the star class, middle-panel to the quasar class and the lower-panel to the galaxy class. A higher density of stars is noticeable at lower latitudes, while more quasars and galaxies clusters at higher magnitudes.}
\label{fig:Data-priors}
\end{figure}

\end{appendix}

\end{document}